\documentclass[11pt,a4paper,aps,preprint,superscriptaddress,nofootinbib]{revtex4-1}
\usepackage{color}
\usepackage[colorlinks=true, pdfstartview=FitV, linkcolor=blue, citecolor=blue, urlcolor=blue]{hyperref}

\usepackage{graphicx,cancel,float}
\usepackage{amssymb,amsmath}
\usepackage{textcomp}
\usepackage{bm,times}
\usepackage{epsfig,epstopdf,multirow}
\usepackage{slashed}

\newcommand{\minigraph}[5][0.25in]{\begin{minipage}{#2}\begin{center}\includegraphics[width=#2]{#5}\\\vspace{#3}\hspace{#1}{\footnotesize #4}\end{center}\end{minipage}}

\def\eqn{eqnarray}
\def\beqn{\begin{\eqn}}
\def\eeqn{\end{\eqn}}
\def\subequ{subequations}
\def\beqs{\begin{\subequ}}
\def\eeqs{\end{\subequ}}

\def\ba{\begin{array}}
\def\ea{\end{array}}

\def\non{\nonumber\\}

\def\hf{\frac{1}{2}}
\def\[{\left[}
\def\]{\right]}
\def\({\left(}
\def\){\right)}

\def\GeV{\rm GeV}

\newcommand{\pslash}{p\hspace{-0.067in}\slash}

\def\gU{\rm U}
\def\gSU{\rm SU}

\def\mM{\mathcal{M}}

\def\mO{\mathcal{O}}

\def\mT{\mathcal{T}}

\def\figureautorefname~#1\null{Fig.\,#1\null}
\def\tableautorefname~#1\null{Tab.\,#1\null}

\def\equationautorefname~#1\null{Eq.\,(#1)\null}
\graphicspath{{fig/}}

\begin{document}

\title{Complementarity of the future $e^+ e^-$ colliders and gravitational waves in the probe of complex singlet extension to the Standard Model}

\author{Ning Chen}
\email{chenning$\_$symmetry@nankai.edu.cn}
\author{Tong Li}
\email{litong@nankai.edu.cn}
\affiliation{School of Physics, Nankai University, Tianjin 300071, China}
\author{Yongcheng Wu}
\email{ycwu@physics.carleton.ca}
\affiliation{Ottawa-Carleton Institute for Physics,
Carleton University, 1125 Colonel By Drive, Ottawa, Ontario K1S 5B6, Canada }
\author{Ligong Bian}
\email{lgbycl@cqu.edu.cn}
\affiliation{Department of Physics,
Chongqing University, Chongqing 401331, China }

\date{\today}

\begin{abstract}
In this work, we study the future probes of the complex singlet extension to the Standard Model (cxSM).
This model is possible to realize a strongly first-order electroweak phase transition (SFOEWPT).
The cxSM naturally provides dark matter (DM) candidate, with or without an exact $\mathbb{Z}_2$ symmetry in the scalar sector.
The benchmark models which can realize the SFOEWPT are selected, and passed to the current observational constraints to the DM candidates, including the relic densities and the direct detection limits set by the latest XENON1T results.
We then calculate the one-loop corrections to the SM-like Higgs boson decays and the precision electroweak parameters due to the cxSM scalar sector.
We perform a global fit to the benchmark models and study the extent to which they can be probed by the future high-energy $e^+ e^-$ colliders, such as CEPC and FCC-ee.
Besides, the gravitational wave (GW) signals generated by the benchmark models are also evaluated.
We further find that the future GW detector, such as LISA, is complementary in probing the benchmark models that are beyond the sensitivity of the future precision tests at the $e^+ e^-$ colliders.
\end{abstract}


\maketitle

\setcounter{page}{2}


\newpage


\section{Introduction}
\label{section:intro}

The observed baryon asymmetry of the Universe (BAU) and the nature of dark matter (DM) are two of the leading puzzles that motivate the new physics beyond the Standard Model (BSM).
One compelling scenario to achieve the BAU is the electroweak baryogenesis (EWBG)~\cite{Kuzmin:1985mm,Cohen:1993nk,Rubakov:1996vz,Funakubo:1996dw,Trodden:1998ym,Morrissey:2012db}.
To preserve the baryon asymmetry generated, a strongly first-order electroweak phase transition (SFOEWPT) is necessary.
It is well-known that the Standard Model (SM) itself cannot realize the SFOEWPT, since the 125 GeV Higgs boson discovered at the LHC~\cite{Aad:2012tfa,Chatrchyan:2012xdj} is too heavy~\cite{Kajantie:1996mn,Rummukainen:1998as,Csikor:1998eu,Aoki:1999fi}.
On the other hand, there is no viable DM candidate in the SM.
To achieve the SFOEWPT and provide possible DM candidate, the SM should be extended.

The simplest realization of the SFOEWPT can be achieved through adding one real scalar singlet to the SM Higgs sector~\cite{OConnell:2006rsp,Ahriche:2007jp,Profumo:2007wc,Curtin:2014jma,Kotwal:2016tex,Hashino:2016xoj,Chao:2017vrq,Kurup:2017dzf,Hashino:2018wee}.
If we impose the $\mathbb{Z}_2$ symmetry under which only the real scalar is odd, this extension can also provide a cold DM candidate since the discrete symmetry forbids the mixing between the neutral doublet and the real singlet.
This scenario admits strongly first order and two-step phase transition in which the singlet scalar acquires a vacuum expectation value (vev) before the electroweak symmetry breaking.
However, in this scenario, the deviations in the $hZZ$ and $hhh$ couplings are induced at loop level.
Thus, no future Higgs factories have the required sensitivities to probe the evidence of such SFOEWPT~\cite{Curtin:2014jma,Huang:2016cjm}.
Besides of the extension of one real scalar singlet, the SFOEWPT can also be realized in the complex scalar extension to the SM (cxSM), as discussed in Refs.~\cite{Barger:2008jx,Jiang:2015cwa, Chiang:2017nmu,Cheng:2018ajh,Grzadkowski:2018nbc,Kanemura:2019kjg,Dev:2019njv}.
DM candidates can naturally arise in the cxSM, in both $\mathbb{Z}_2$ symmetric and $\mathbb{Z}_2$ breaking scenarios~\footnote{See Refs.~\cite{Wu:2016mbe,Gross:2017dan,Ishiwata:2018sdi,Huitu:2018gbc,Cline:2019okt} for the DM phenomena in the cxSM. See Ref~\cite{Karamitros:2019ewv} for a cancellation of direct detection for pseudo Nambu-Goldstone DM. }.
Hence, the cxSM is appealing to address both the SFOEWPT issue and the DM candidate at one shot.
The next question is whether the cxSM with SFOEWPT and DM candidate can be explored by the future experiments.

Direct searches for the extended scalar sector beyond the SM have been carried out in the Large Hadron Collider (LHC) experiments~\cite{Coleppa:2014hxa,Coleppa:2014cca,Chen:2014dma,Craig:2015jba,Li:2015lra,Kling:2015uba,Chen:2015fca,Craig:2016ygr,Kling:2018xud,Djouadi:2019cbm}.
No signal has been reported so far.
Due to the small mixing effects of the ${\rm SU}(2)_L$ singlets, it is expected that the direct search for the scalars from the complex singlet is very challenging at the LHC~\cite{Kotwal:2016tex,Huang:2017jws,Li:2019tfd}~\footnote{Ref.~\cite{Basler:2019nas} shows that the possible interference effects via the $t \bar t$ and $hh$ final states from the cxSM are suppressed comparing to the two-Higgs doublet model.}.
Complementary to the direct searches, the precision measurements of the Higgs boson properties could shed light on the underlying new physics.
Several well-known proposals have been made to build the next-generational Higgs factory, such as the Circular Electron Positron Collider (CEPC) in China~\cite{CEPC-SPPCStudyGroup:2015csa,CEPCStudyGroup:2018ghi}, the electron-positron stage of the Future Circular Collider (FCC-ee) at CERN~\cite{Gomez-Ceballos:2013zzn}, and the International Linear Collider (ILC) in Japan~\cite{Baer:2013cma,Bambade:2019fyw}.
Each facility is proposed to run at $\sqrt{s}=240-250\,\GeV$ to produce $10^5 - 10^6$ SM-like Higgs bosons, aiming to reach sub-percentage precision measurement of its couplings.
Besides, they will also run at the $Z$-pole to improve the precisions on the measurement of SM parameters by a factor of $20-200$ over the results from Large Electron Positron (LEP) Collider~\cite{ALEPH:2005ab}.
With such incredible improvements in the precision measurements, a number of studies have been carried out to look for the BSM effects through both tree-level and one-loop corrections to the productions~~\cite{Englert:2013tya,Craig:2013xia,McCullough:2013rea,Curtin:2014jma} and decays~\cite{Huang:2016cjm,Gu:2017ckc,Li:2017daq,Kanemura:2018yai,Chen:2018shg} of Higgs boson at the future $e^+ e^-$ colliders~\footnote{Recently, the computation of the one-loop corrected Higgs boson couplings in the extended Higgs sector was provided in the package of {\tt H-COUP}~\cite{Kanemura:2017gbi,Kanemura:2019slf}.}.

Even with the precision measurements of the Higgs boson properties at the future $e^+ e^-$ colliders, one might encounter the so-called ``nightmare scenario'' where model points are inaccessible at the colliders~\cite{Curtin:2014jma,Huang:2016cjm}.
Initiated by the detection of the GWs from a binary black hole merger by LIGO/VIRGO~\cite{Abbott:2016blz}, the detection of the gravitational waves (GW) may provide a complementary probe of models that can achieve the SFOEWPT~\cite{Huang:2016odd,Chen:2017cyc,Baldes:2018nel,Chiang:2018gsn,Baldes:2018emh,Ellis:2018mja,Madge:2018gfl,Beniwal:2018hyi,Shajiee:2018jdq,Alves:2018jsw,Brdar:2019fur,Alanne:2019bsm}.
If a SFOEWPT occurred in the early Universe, the bubble collisions and the damping of plasma inhomogeneities are expected to generate a stochastic background of GWs.
For an electroweak phase transition, the peak frequencies of the GW spectrum happen around $\mO(10^{-4})-\mO(10^{1})$ Hz, which are potentially within the reach of future space-based GW interferometers, such as LISA~\cite{AmaroSeoane:2012km,AmaroSeoane:2012je,Caprini:2015zlo,Audley:2017drz} and its successor BBO~\cite{Cutler:2009qv}, Taiji~\cite{Guo:2018npi}, Tianqin~\cite{Luo:2015ght}, Decigo~\cite{Kawamura:2011zz}, and beyond~\cite{Crowder:2005nr,Corbin:2005ny,Baker:2019pnp}.

In this work, we study the future experimental tests of the cxSM, including the precision measurements at $e^+ e^-$ colliders and the sensitivity of GW signal.
The scenario of the cxSM is possible to achieve the SFOEWPT and provide viable cold DM candidate.
We perform the global fit to the full parameter space by requiring the conditions for a SFOEWPT and focus on the possibility of GW and future Higgs factories as a probe of SFOEWPT.
The constraints from DM relic density as well as the lower limits on the spin-independent (SI) DM-nucleon scattering cross section set by the latest direct detection (DD) experiments will be imposed on the model parameters.
The corresponding GW spectra due to the SFOEWPT will be evaluated by including the contributions from bubble wall collisions, the sound waves, and the magnetohydrodynamic (MHD) turbulence.
We further estimate the signal-to-noise ratio (SNR) of the model points that can achieve the SFOEWPT and find their sensitivity in the future LISA interferometer.
We follow the Ref.~\cite{Chen:2018shg} to perform the combined $\chi^2$ fit to the precision measurements of the electroweak parameters and the one-loop corrections to the Higgs boson decays.

The rest of this paper is organized as follows.
In Sec.~\ref{section:cxSM}, we review the framework of the cxSM, and list the corresponding mass spectra and the relevant cubic Higgs self-couplings.
In Sec.~\ref{section:constraints}, we impose the theoretical constraints on the cxSM potential, as well as the constraints on the DM candidate in the cxSM.
In Sec.~\ref{section:SFOEWPT}, we discuss the SFOEWPT in the cxSM by making use of the finite-temperature effective potential, as well as the GW signals.
We also give the one-loop corrections to the SM-like Higgs boson couplings, and the electroweak precision observables changed by the cxSM.
In Sec.~\ref{section:results}, the benchmark models that can realize the SFOEWPT will be used for the precision tests at the future $e^+ e^-$ colliders.
We show our numerical results for the parameter space that can be probed by future experiments.
The conclusion is given in Sec.~\ref{section:conclusion}.


\section{The complex singlet extension to the SM}
\label{section:cxSM}

\subsection{The Higgs potential and global symmetries}

We extend the SM Higgs sector by introducing a complex scalar singlet $\mathbb{S}$ of the $\gSU(2)_L$.
The most general scalar potential in this extension is expressed as~\cite{Barger:2008jx}
\beqn\label{eq:VcxSM}
V(\Phi\,, \mathbb{S})&=& \mu^2 | \Phi |^2 +  \lambda | \Phi |^4 + \frac{\delta_2 }{2} |\Phi|^2 | \mathbb{S} |^2  +  \frac{b_2}{2} | \mathbb{S} |^2 +  \frac{d_2}{4} | \mathbb{S} |^4  \non
&+& \Big( \frac{ \delta_1  }{4 } |\Phi |^2 \mathbb{S} +  \frac{ \delta_3   }{4 } |\Phi |^2 \mathbb{S}^2 + c.c. \Big)\non
&+& \Big( a_1 \mathbb{S}  +  \frac{ b_1 }{4} \mathbb{S}^2 +   \frac{ c_1  }{6} \mathbb{S}^3  +  \frac{ c_2 }{6} \mathbb{S} |\mathbb{S}|^2  + \frac{  d_1  }{8} \mathbb{S}^4 + \frac{  d_3    }{8} \mathbb{S}^2 |\mathbb{S}|^2 + c.c. \Big)\,,
\eeqn
where $\Phi$ is the $\gSU(2)_L$ Higgs doublet breaking the electroweak symmetry.
The parameters in the first line of Eq.~\eqref{eq:VcxSM} are real, and the other parameters in the second and third lines of Eq.~\eqref{eq:VcxSM} are generally complex.
Two possible global symmetries can be imposed to the above Higgs potential:
\begin{itemize}

\item A discrete $\mathbb{Z}_2$ symmetry of $\mathbb{S} \to - \mathbb{S}$ can be imposed to eliminate all terms with odd powers of $\mathbb{S}$, which include $\delta_1$, $a_1$, $c_{1\,,2}$\,.

\item A global $\gU(1)$ symmetry of $\mathbb{S} \to e^{i\alpha} \mathbb{S}$ eliminates all terms with complex coefficients ($\delta_{1\,,3}$, $a_1$, $b_1$, $c_{1\,,2}$, $d_{1\,,3}$)\, .

\end{itemize}
If the complex scalar field $\mathbb{S}$ does not obtain a zero-temperature vev, the discrete $\mathbb{Z}_2$ symmetry has to be introduced to stabilize the scalar singlet and enable DM candidate(s).
Under a further global $\gU(1)$ symmetry, this cxSM model yields two degenerate stable DM particles (the two components in $\mathbb{S}$).
This case with only the terms in the first line of Eq.~(\ref{eq:VcxSM}) is very similar to the real singlet model.
By including one $\gU(1)$ breaking term, for instance the $b_1$ term, the real and imaginary parts of $\mathbb{S}$ are still stable but not identical anymore.
Below we refer this more general case as the $\mathbb{Z}_2$ symmetric scenario with the following scalar potential
\beqn
V(\Phi\,, \mathbb{S})_{\mathbb{Z}_2}&=& \mu^2 | \Phi |^2 +  \lambda | \Phi |^4 + \frac{\delta_2 }{2} |\Phi|^2 | \mathbb{S} |^2  +  \frac{b_2}{2} | \mathbb{S} |^2 +  \frac{d_2}{4} | \mathbb{S} |^4  +\frac{ b_1 }{4}\Big( \mathbb{S}^2 + c.c. \Big)\, .
\label{Z2}
\eeqn
On the other hand, if the $\mathbb{S}$ field acquires a zero-temperature vev and thus the real component of $\mathbb{S}$ mixes with the neutral Higgs of $\Phi$, the $\gU(1)$ and $\mathbb{Z}_2$ symmetries are both spontaneously broken by the singlet vev and the Goldstone boson from the imaginary part of $\mathbb{S}$ is stable but massless.
To provide a viable DM candidate, a soft breaking of the global $\gU(1)$ symmetry is introduced to generate a mass for it.
The $\gU(1)$ breaking requires that one or more terms in the second and third lines of Eq.~(\ref{eq:VcxSM}) does not vanish.
We demand that $b_1\neq 0$ as well, and the $\gU(1)$ symmetry is both spontaneously and softly broken.
Now the spontaneously broken $\mathbb{Z}_2$ symmetry may lead to the cosmological domain wall problem~\cite{Zeldovich:1974uw,Friedland:2002qs}.
To solve this problem, one can further introduce one or more of $\delta_1, a_1, c_{1,2}$ terms to explicitly break the $\mathbb{Z}_2$ symmetry.
We consider the following potential with a non-vanishing $a_1$ as in Ref.~\cite{Barger:2008jx}
\beqn
V(\Phi\,, \mathbb{S})_{\cancel{\mathbb{Z}_2}}&=&  \mu^2 |\Phi|^2 +  \lambda |\Phi|^4 + \frac{\delta_2}{2} |\Phi|^2 | \mathbb{S} |^2  + \frac{b_2}{2} | \mathbb{S} |^2 + \frac{d_2}{4} | \mathbb{S} |^4 \non
&+& \Big(a_1  \mathbb{S} +  \frac{b_1}{4}  \mathbb{S}^2 + c.c. \Big)\,.
\label{Z2break}
\eeqn
We refer the above potential as the $\mathbb{Z}_2$ breaking scenario below.
One should keep in mind that, although we follow the choices of Ref.~\cite{Barger:2008jx} in the rest of this paper, the scalar potential for achieving the above purposes is not unique.

\subsection{The $\mathbb{Z}_2$ symmetric scenario}

To minimize the scalar potential, we represent the complex scalar singlet as $\mathbb{S} = \frac{1}{ \sqrt{2} } (S + i A)$ and the Higgs doublet as $\Phi= (0\,,h/ \sqrt{2})^T$.
In the $\mathbb{Z}_2$ symmetric case, we only have the SM Higgs doublet developing a vev ($v$) and the $a_1$ term is vanishing.
From Eq.~\eqref{Z2}, the field-dependent scalar potential at the tree level becomes
\beqn
V_0(h\,,S\,,A)&=& \frac{ \mu^2 }{2} h^2 + \frac{\lambda}{4} h^4 + \frac{ \delta_2 }{8} h^2 (S^2 + A^2 ) \non
&+&  \frac{1}{4} ( b_1 + b_2 ) S^2  + \frac{1}{4} ( b_2 - b_1 ) A^2   + \frac{d_2}{ 16} (S^2 + A^2 )^2\,.
\eeqn
By minimizing the potential, one arrives at the following condition
\beqn
0&=&\frac{\partial V_0}{\partial h}\Big|_{h=v\,, S=0\,,A=0}= \mu^2 v + \lambda v^3  \Rightarrow \mu^2 = -\lambda v^2 \,.
\eeqn
The mass spectrum is obtained as follows
\beqs
\beqn
M_A^2&=& \frac{\partial^2 V_0}{\partial A^2} \Big|_{h=v\,, S= 0\,, A=0}= \frac{1}{4} \delta_2 v^2 - \frac{1}{2} (b_1 - b_2 )  \,,\\
M_h^2 &=& \frac{\partial^2 V_0}{ \partial h^2} \Big|_{h=v\,, S= 0\,,A=0} = 2 \lambda v^2\,,\\
M_S^2&=& \frac{\partial^2 V_0}{ \partial S^2} \Big|_{h=v\,, S= 0\,,A=0} =  \frac{1}{4} \delta_2 v^2 +  \hf (b_1 + b_2 )  \,.
\eeqn
\eeqs
With the exact $\mathbb{Z}_2$ symmetry, $h$ and $S$ do not mix.
Both $S$ and $A$ are stable and regarded as the DM candidates in our following discussions.
Altogether, the parameters in the generic basis and the physical basis are
\beqs
\beqn
\textrm{generic basis}&:& \mu^2\,,\quad \lambda\,, \quad \delta_2\,, \quad b_1 \,,\quad b_2\,,\quad d_2 \,;\\
\textrm{physical basis}&:& M_{h\,,S\,,A}\,,\quad v\,, \quad \delta_2\,,\quad d_2 \,,
\eeqn
\eeqs
with the fixed inputs as $M_h=125\,\GeV$ and $v\approx 246\,\GeV$.
The ranges of remaining parameters we take for the scan are
\beqn
&& 65\,{\rm GeV} \leq M_S \leq 2000\,{\rm GeV}\,,\quad   65\,{\rm GeV} \leq M_A \leq 2000\,{\rm GeV}\, , \non
&& 0 \leq d_2 \leq 20 \,,\quad  -20 \leq \delta_2 \leq 20\,.
\label{scanrange:Z2}
\eeqn
%

\subsection{The $\mathbb{Z}_2$ breaking scenario}

In the $\mathbb{Z}_2$ breaking scenario, the field-dependent scalar potential at the tree level reads
\beqn
V_0(h\,,S\,,A)&=& \hf \mu^2 h^2 + \frac{\lambda}{4} h^4 + \frac{ \delta_2 }{8} h^2 (S^2+A^2)  + \sqrt{2} a_1 S + \frac{b_1 + b_2}{4} S^2  + \frac{ d_2 }{16}  S^4 \nonumber \\
&+& \frac{-b_1 + b_2}{4} A^2 + \frac{ d_2 }{16}  A^4 + \frac{ d_2 }{8}  S^2 A^2\,.
\eeqn
Both $h$ and $S$ obtain vevs in this case.
The corresponding minimization conditions are
\beqs
\beqn
0&=&\frac{\partial V_0}{\partial h}\Big|_{h=v\,, S=v_s\,,A=0}= \mu^2 v + \lambda v^3 + \frac{\delta_2}{4} v v_s^2\non
& \Rightarrow& \mu^2 = -\lambda v^2 - \frac{\delta_2}{4} v_s^2 \,,\label{eq:V0mincond1}\\
0&=& \frac{\partial V_0}{ \partial S}\Big|_{h=v\,, S=v_s\,,A=0}= \frac{\delta_2}{4} v^2 v_s + \sqrt{2 } a_1 + \frac{ b_1 + b_2}{2} v_s + \frac{d_2}{4} v_s^3 \non
&\Rightarrow & b_1 + b_2 = - 2 \sqrt{2} \frac{ a_1 }{ v_s }  - \frac{ \delta_2 }{2} v^2 - \frac{d_2}{2} v_s^2\, .\label{eq:V0mincond2}
\eeqn
\eeqs
The mass spectrum of the scalars for the $\mathbb{Z}_2$ breaking scenario is obtained as follows
\beqs
\beqn
M_A^2&=& \frac{\partial^2 V_0}{\partial A^2} \Big|_{h=v\,, S= v_s\,, A=0}= - b_1 - \sqrt{2} \frac{ a_1 }{ v_s} \,,\\
\mM^2&=&  \left(  \ba{cc}
\mu_h^2 & \mu_{hs}^2  \\
\mu_{hs}^2 &  \mu_s^2   \ea \right)  \,, \\
\mu_h^2 &=& \frac{\partial^2 V_0}{ \partial h^2} \Big|_{h=v\,, S= v_s\,, A=0} = 2 \lambda v^2\,,\\
\mu_s^2&=& \frac{\partial^2 V_0}{ \partial S^2} \Big|_{h=v\,, S= v_s\,, A=0} = \frac{d_2}{2} v_s^2 - \sqrt{2}\frac{a_1}{v_s} \,,\\
\mu_{hs}^2&=& \frac{\partial^2 V_0}{ \partial h \partial S} \Big|_{h=v\,, S= v_s\,, A=0}=  \frac{ \delta_2 }{2}  v v_s \,.
\eeqn
\eeqs
The mass eigenstates after diagonalizing the CP-even scalars are
\beqn
\left( \ba{c} h_1  \\ h_2  \ea  \right) &=&\left(
\ba{cc}
\cos\theta & \sin\theta   \\
-\sin\theta & \cos\theta   \\
\ea \right) \left( \ba{c} h  \\ S \ea  \right)\,,
\eeqn
with the masses of $h_1$ and $h_2$ being $M_1$ and $M_2$, respectively.
The CP-odd component $A$ will not develop a vev and will be treated as the DM candidate for the later discussion.

In terms of the mass eigenstates, our parameter inputs can be traded into the CP-even scalar masses and the mixing angle as
\beqs\label{eqs:SelfCoup}
\beqn
\lambda&=&\frac{1}{2 v^2} \Big( \cos^2\theta M_1^2 + \sin^2\theta M_2^2  \Big) \,,\\
  \frac{d_2}{2} &=& \frac{1}{v_s^2 } \Big( \sin^2\theta M_1^2 + \cos^2\theta M_2^2 + \sqrt{2}\frac{a_1}{v_s} \Big) \,,\\
 \delta_2 &=& \frac{2}{ v v_s} ( M_1^2 - M_2^2  ) \cos\theta \sin\theta \,.
\eeqn
\eeqs
Altogether, the parameters in the generic basis and the physical basis are
\beqs
\beqn
\textrm{generic basis}&:& \mu^2\,,\quad \lambda\,, \quad \delta_2\,, \quad a_1\,, \quad b_1 \,,\quad b_2\,,\quad d_2 \,,\\
\textrm{physical basis}&:& M_{1\,,2\,,A}\,,\quad \theta\,,\quad v\,,\quad v_s\,,\quad a_1 \,,
\eeqn
\eeqs
with the fixed inputs of $M_1=125\,\GeV$ and $v\approx 246\,\GeV$.
Below, we scan the physical parameters in the following ranges
\beqn\label{eq:scanrange}
&& 0 \leq v_s\leq 150 \ {\rm GeV}, \ \ 65 \ {\rm GeV} \leq M_2 \leq 150 \ {\rm GeV}\, , \ \ 65 \ {\rm GeV} \leq M_A \leq 2000 \ {\rm GeV}\, , \non
&& 0\leq \theta \leq 0.5, \ \  -(100\,{\rm GeV})^3\leq  a_1 \leq (100\,{\rm GeV})^3\,.
\label{scanrange:Z2break}
\eeqn
In both scenarios, the mass ranges of $M_{S,2}$ taken in our study make it easier to achieve the two-step phase transition.
It turns out that the corresponding choice of mass ranges guarantees that the EWSB vacuum of $\langle h \rangle \neq 0$ is the global minimum, as comparing to the electroweak symmetric vacuum~\cite{Bian:2018mkl,Cheng:2018ajh}.

\subsection{The Higgs self-couplings}

In the $\mathbb{Z}_2$ symmetric scenario, the relevant cubic and quartic Higgs self-couplings are listed below
\begin{subequations}
  \label{equ:Z2CCoupling}
\begin{align}
 \lambda_{hhh}&= \frac{M_h^2}{2v} \,, \\
 \lambda_{hSS}&= \lambda_{hAA} = \frac{1}{4} \delta_2 v\, , \\
 \lambda_{hhhh}&= \frac{M_h^2}{8v^2} \, .
\end{align}
\end{subequations}
Here and below we define the cubic and quartic self-couplings as the coefficients of scalar fields in the Lagrangian. The cubic and quartic Higgs self-couplings of the SM-like Higgs bosons are the same as those in the SM case, while the other two Higgs couplings $\lambda_{hSS}$ and $\lambda_{hAA}$ are relevant for the Higgs boson self-energy corrections at the one-loop level.

In the $\mathbb{Z}_2$ breaking scenario, the relevant cubic Higgs and quartic self-couplings in the physical basis can be expressed as follows
\begin{subequations}
  \label{equ:Z2VCoupling}
\begin{align}
  \lambda_{111} &= \frac{s_\theta^3(\sqrt{2}a_1+M_1^2v_s)}{2v_s^2} + \frac{M_1^2c_\theta^3}{2v}\, , \\
  \lambda_{112} &= \frac{s_{2\theta}}{4v v_s^2}\left(3\sqrt{2}a_1vs_\theta+v_s(2M_1^2+M_2^2)(vs_\theta-v_sc_\theta)\right)\, , \\
  \lambda_{122} &= \frac{s_{2\theta}}{4v v_s^2}\left(3\sqrt{2}a_1vc_\theta+v_s(M_1^2+2M_2^2)(vc_\theta+v_ss_\theta)\right)\, , \\
  \lambda_{1AA} &= \frac{s_\theta}{2v_s^2}\left(\sqrt{2}a_1+ M_1^2v_s\right)\, , \\
  \lambda_{2AA} &= \frac{c_\theta}{2v_s^2}\left(\sqrt{2}a_1+ M_2^2v_s\right)\, , \\
  \lambda_{222} &= \frac{\sqrt{2}a_1c_\theta^3}{2v_s^2}+\frac{M_2^2}{2vv_s}\left(vc_\theta^3-v_ss_\theta^3\right)\, , \\
  \lambda_{1111} &= {c_\theta^6 M_1^2+c_\theta^4 s_\theta^2 M_2^2\over 8v^2} + {c_\theta^3 s_\theta^3 (M_1^2-M_2^2)\over 4vv_s} + {s_\theta^4(\sqrt{2}a_1+c_\theta^2 M_2^2 v_s+s_\theta^2M_1^2 v_s)\over 8v_s^3}\, .
\end{align}
\end{subequations}
We use the subscripts $1, 2, A$ to manifest the couplings among $h_1, h_2, A$ scalars, respectively. The cubic and quartic self-couplings recover the SM couplings when the mixing angle $\theta\to 0$.
We define the deviations of the cubic and quartic self-couplings of the SM-like Higgs as $\delta \kappa_3\equiv\lambda_{111}/\lambda_{hhh}^{\rm SM}-1$ and $\delta \kappa_4\equiv \lambda_{1111}/\lambda_{hhhh}^{\rm SM}-1$.
The correlation between them guarantees the tree-level driven SFOEWPT.


\section{Constraints}
\label{section:constraints}

\subsection{Unitarity and stability}

In order to have a well-defined Higgs potential, a set of theoretical constraints should be taken into account.
The Lee-Quigg-Thacker unitarity bound~\cite{Lee:1977yc,Lee:1977eg} should be imposed so that the quartic couplings are not too large.
In both $\mathbb{Z}_2$ symmetric and $\mathbb{Z}_2$ breaking scenarios, the quartic terms of the Higgs potential are the following
\beqn
V_0&\sim& \lambda \Big( \hf h^2 + \hf (\pi^0)^2 + \pi^+ \pi^-  \Big)^2 + \frac{\delta_2 }{4} (S^2 + A^2 ) ( \hf h^2 + \hf (\pi^0)^2 + \pi^+ \pi^- ) \non
&+& \frac{d_2}{16 } (S^2 + A^2)^2\,.
\eeqn
By taking the neutral states of $| \pi^+ \pi^- \rangle $, $\frac{1}{\sqrt{2} } | \pi^0 \pi^0 \rangle$, $\frac{1}{ \sqrt{2} } | h h \rangle$, $\frac{1}{ \sqrt{2} } | SS \rangle$, and $\frac{1}{ \sqrt{2} } | AA \rangle$, the $s$-wave matrix reads
\beqn\label{eq:a0matrix}
a_0&=& \frac{1}{16\pi} \left(
\ba{ccccc}
4\lambda & \sqrt{2} \lambda  & \sqrt{2} \lambda & \frac{\delta_2}{2 \sqrt{2} }  & \frac{\delta_2}{2 \sqrt{2} }   \\
\sqrt{2} \lambda & 3 \lambda  & \lambda  & \frac{\delta_2}{4} & \frac{\delta_2}{4}  \\
 \sqrt{2} \lambda & \lambda  & 3 \lambda  & \frac{\delta_2}{4} & \frac{\delta_2}{4}  \\
 \frac{\delta_2}{2 \sqrt{2} }  &  \frac{\delta_2}{4}  & \frac{\delta_2}{4}   &  \frac{3 d_2}{4}  &   \frac{d_2}{4}   \\
 \frac{\delta_2}{2 \sqrt{2} }  & \frac{\delta_2}{4}   & \frac{\delta_2}{4}   &   \frac{d_2}{4}  &  \frac{3 d_2}{4}  \\
\ea \right)\,.
\eeqn
The $s$-wave unitarity conditions are imposed such that $|\tilde a_0^i|\leq 1$, with $\tilde a_0^i$ being all eigenvalues of matrix $a_0$ above.
By using the relations in Eqs.~\eqref{eqs:SelfCoup}, the perturbative unitarity condition can impose bounds to the Higgs boson masses and mixings.
In addition, one should impose the following tree-level stability conditions so that the scalar potential is bounded from below at the large field values
\beqn
&& \lambda > 0 \,, \qquad d_2 > 0 \,, \qquad \lambda d_2 > \delta_2^2\,.
\eeqn
Here, the last term is necessary for $\delta_2<0$.

\subsection{The global minimum}

In terms of the classical fields, there may be three different configurations for the symmetry breaking:
\beqn
O&:& h\to 0\,,\quad S \to 0 \,; \non
A&:& h \to 0 \,,\quad S \to v_s \,; \non
B&:& h \to v \,,\quad S \to 0 \ (v_s)\,,
\eeqn
for the $\mathbb{Z}_2$ symmetric (breaking) cases, respectively.
As the temperature cools down, the symmetry breaking may occur either by one step via $O \to B$, or by two steps via $O \to A\to B$.
The one-step phase transition occurs if the configuration-$B$ is the only possible Higgs potential minimum, and the two-step phase transition occurs if both configure-$A$ and configuration-$B$ coexist as the Higgs potential minimum.

The EWSB vacuum solution of $B$ should be the lowest one of the scalar potential, while the origin point of $O$ should be the highest one.
The vacuum configurations of $A$ and $B$ are obtained by solving the following cubic equations
\beqs\label{eqs:dVT_highT}
\beqn
A&:&\frac{\partial V_0}{\partial S}\Big|_{h=0\,, S=v_s\,,A=0}=0\,,\\
B&:& \frac{\partial V_0}{\partial h}\Big|_{h=v\,, S=0\,,A=0}=0 \quad {\rm \mathbb{Z}_2 \ symmetric }\,,\\
B&:& \frac{\partial V_0}{\partial h}\Big|_{h=v\,, S=v_s\,,A=0}=0 \,, \frac{\partial V_0}{\partial S}\Big|_{h=v\,, S=v_s\,,A=0}=0 \quad {\rm  \mathbb{Z}_2 \ breaking }\,.
\eeqn
\eeqs
The numerical solutions are then fed into $V_0(A)$ and $V_0(B)$, and the global minimum condition $V_0(B) \leq V_0(A)$ will be imposed.

\subsection{The constraints on the DM candidates}

\begin{figure}
\begin{center}
\minigraph{12cm}{-0.05in}{(a)}{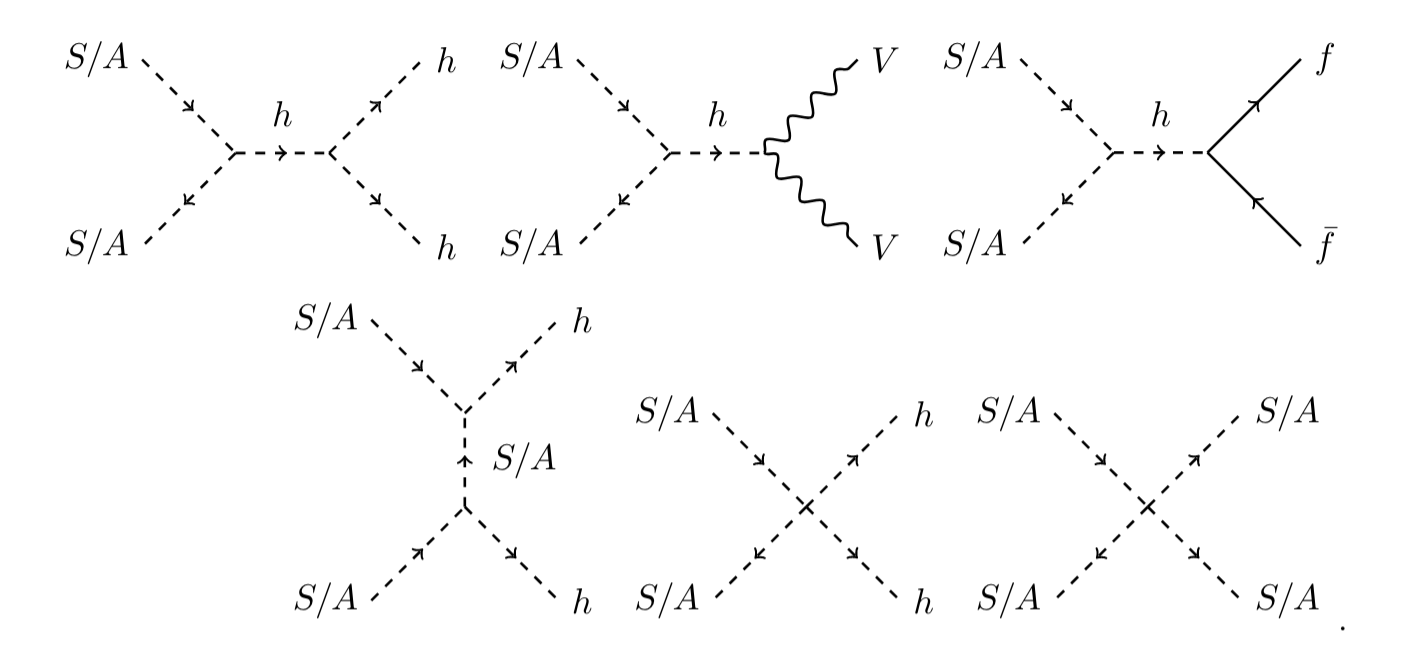}
\minigraph{12cm}{-0.05in}{(b)}{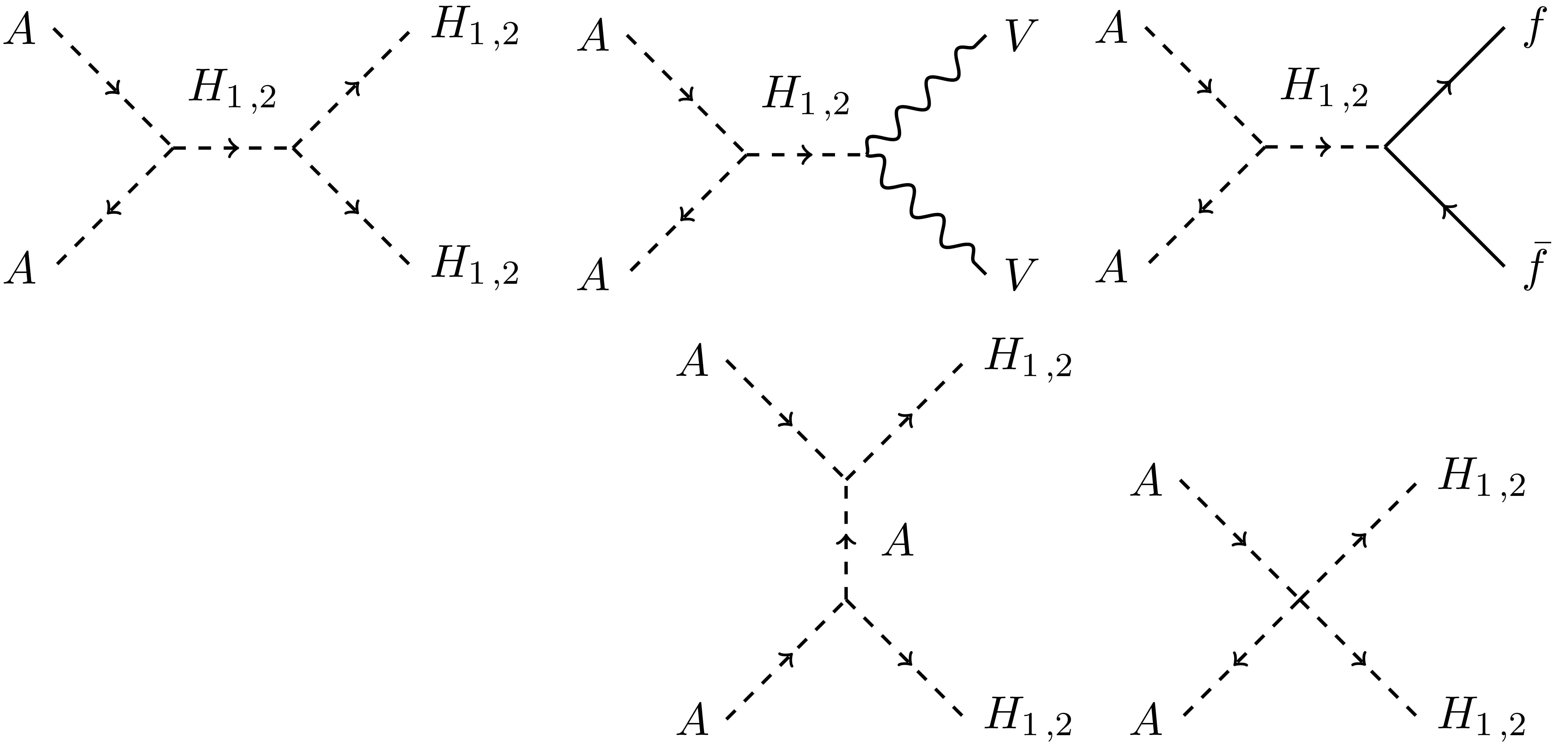}
\end{center}
\caption{
The DM annihilation processes for the $\mathbb{Z}_2$ symmetric scenario (a) and the $\mathbb{Z}_2$ breaking scenario (b).
}
\label{fig:DMann}
\end{figure}

In the $\mathbb{Z}_2$ symmetric scenario with $v_s=0$, both $S$ and $A$ are regarded as stable particles. They both contribute to the total relic abundance depending on the parameter $d_2$ and their mass splitting~\cite{Barger:2008jx}.
In the $\mathbb{Z}_2$ breaking scenario, only the CP-odd scalar $A$ becomes the DM candidate.
The annihilation processes that contribute to the DM relic density in the two cases are shown in Fig.~\ref{fig:DMann}.
The relic density typically exhibits one (two) dip(s) with the DM mass being around $M_h/2$ ($M_1/2$ or $M_2/2$), due to the enhancement of the annihilation cross section near the $h$ ($h_1, h_2$) resonance(s) in the $\mathbb{Z}_2$ symmetric (breaking) scenario.

Several ongoing DD experiments are looking for DM scattering off atomic nuclei, including XENON1T~\cite{Aprile:2018dbl} and PandaX-II~\cite{Cui:2017nnn}.
No conclusive observation has been reported so far.
For the DM mass range of $\mO(10) - \mO(10^3)\,\GeV$, XENON1T has set the most stringent lower limit on the SI DM-nucleon scattering cross section as $\sigma_{\rm SI} \lesssim 10^{-46} - 10^{-44}\,{\rm cm}^2$~\cite{Aprile:2017iyp,Aprile:2018dbl}.
For the $\mathbb{Z}_2$ symmetric case, the SI scattering processes are mediated only by the SM-like Higgs boson $h$; while for the $\mathbb{Z}_2$ breaking case, the SI scattering processes are mediated by two CP-even scalars of $h_{1\,,2}$.
The corresponding cross sections are given by~\cite{Barger:2010yn,Gonderinger:2012rd,Jiang:2015cwa}
\beqs
\beqn
\sigma_{\rm SI}(\mathbb{Z}_2)&=& \frac{m_p^4}{ 2\pi v^2 } \frac{1}{ (m_p + M_i )^2} \Big( \frac{\lambda_{hii} }{ M_h^2 } \Big)^2 \Big( f_{Tu}^{(p)}  + f_{Td}^{(p)}  + f_{Ts}^{(p)}  + \frac{2}{9} f_{TG}^{(p)}    \Big)^2\,, \ i=S, A \,,\label{eq:sigmaSI_Z2}\\
\sigma_{\rm SI} (\cancel{\mathbb{Z}_2})&=& \frac{ m_p^4 }{ 2\pi v^2 ( m_p + M_A)^2} \Big(  \frac{ \lambda_{1AA } \cos\theta}{M_1^2 } -  \frac{ \lambda_{2 AA} \sin\theta }{M_2^2}  \Big)^2\non
&&\times \Big( f_{Tu}^{(p)}  + f_{Td}^{(p)}  + f_{Ts}^{(p)}  + \frac{2}{9} f_{TG}^{(p)} \Big)^2\,,\label{eq:sigmaSI_Z2b}
\eeqn
\eeqs
with the nucleon form factors of $f_{Tu}^{(p)}, f_{Td}^{(p)}, f_{Ts}^{(p)}$ and $f_{TG}^{(p)} = 1- \sum_{q=u\,,d\,,s} f_{Tq}^{(p)}$.
The possible cancellation between the $h_1$ and $h_2$ diagrams, as indicated by Eq.~\eqref{eq:sigmaSI_Z2b}, leads to further suppressed scattering cross section for the $\mathbb{Z}_2$ breaking scenario compared with the $\mathbb{Z}_2$ symmetric scenario.
Note that if $a_1=0$ in this case, the two diagrams exactly cancel each other for any sets of model parameters due to the pseudo-Goldstone nature of the DM particle $A$ here~\cite{Gross:2017dan}.

\begin{figure}
\centering
\minigraph{7cm}{-0.05in}{(a)}{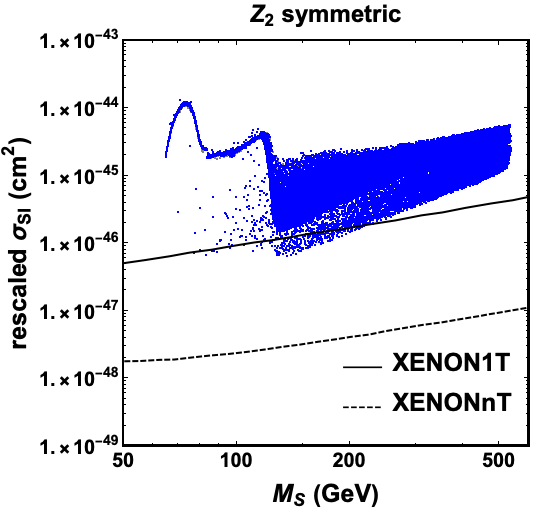}
\minigraph{7cm}{-0.05in}{(b)}{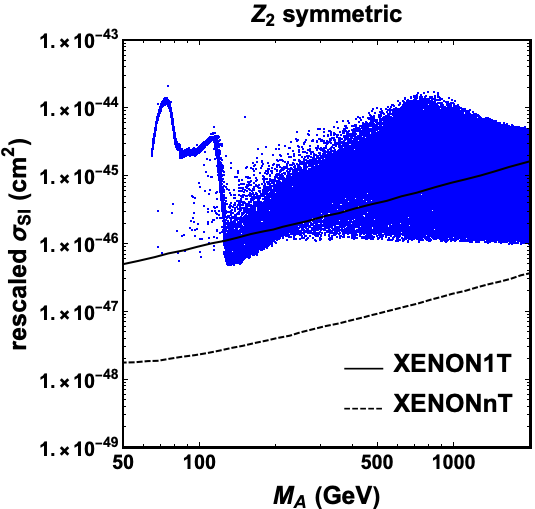}\\
\minigraph{7cm}{-0.05in}{(c)}{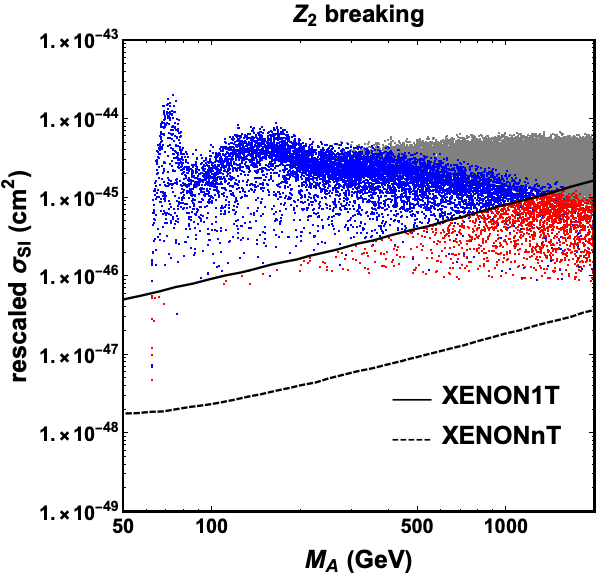}
\caption{
The rescaled SI cross sections of the DM candidate for the $\mathbb{Z}_2$ symmetric scenario (a, b) and the $\mathbb{Z}_2$ breaking scenario (c).
The grey points are those oversaturate the relic density.
The blue points satisfy the relic density requirement.
The red points are those satisfy both the relic density requirement and the current direct detection limit by the XENON1T (solide line), and pass the SFOEWPT criterion following the approach in Sec.~\ref{section:SFOEWPT}.
The future projected limit by the XENONnT is also displayed (dashed line).
}
\label{fig:cxSMDMdd}
\end{figure}

In practice, we first produce the {\tt CalcHEP}~\cite{Belyaev:2012qa} model files by implementing the cxSM model parameters and interactions in {\tt FeynRules}~\cite{Christensen:2008py} .
The model files are then fed to {\tt MicrOMEGAs}~\cite{Belanger:2018mqt} to calculate the DM relic density for the cxSM model (denoted as $\Omega_{\rm cxSM} h^2$) and the SI scattering cross section $\sigma_{\rm SI}$.
The above quark/gluon-nucleon form factors are taken as the default values in {\tt MicrOMEGAs}.
The current measurements of the cold DM relic density are given as $\Omega_{\rm DM}h^2 = 0.1138\pm 0.0045$ (WMAP)~\cite{Hinshaw:2012aka} or $\Omega_{\rm DM}h^2 = 0.1196\pm 0.0031$ (Planck)~\cite{Ade:2013zuv}.
After the scan of parameter spaces in Eqs.~(\ref{scanrange:Z2}) and (\ref{scanrange:Z2break}) by imposing the above theoretical constraints, the survived points that oversaturate the relic density are further rejected.
For those points that undersatuarate the relic density, we rescale the SI cross section by
\beqn
&&\sigma_{\rm SI} ({\rm rescaled}) = \sigma_{\rm SI} \cdot \frac{ \Omega_{\rm cxSM} h^2}{ \Omega_{\rm DM} h^2 } \,,
\eeqn
and compare with the latest limit from the XENON1T~\cite{Aprile:2018dbl}.
In Fig.~\ref{fig:cxSMDMdd}, the rescaled SI cross sections of model points are evaluated and the current limit set by the XENON1T experiment is added as reference.
The model points satisfying both the relic density constraint $\Omega_{\rm cxSM} h^2 < \Omega_{\rm DM} h^2$ and the current XENON1T DD limit and passing the SFOEWPT criterion are marked in red. The $hh$ channels are kinematically allowed for $M_{S,A}$ heavier than 125 GeV.
These additional channels reduce the relic density and the parameter space opens up near above 125 GeV.
As seen from the plots (a) and (b) of Fig.~\ref{fig:cxSMDMdd}, we do not find any parameter choices viable for SFOEWPT and satisfying DM constraint in the $\mathbb{Z}_2$ symmetric scenario.
They are not compatible in this case as concluded in Ref.~\cite{Chiang:2017nmu}.
Thus, in the numerical study below, we will mainly focus on the $\mathbb{Z}_2$ breaking scenario and apply this constraint on the benchmark points for later studies.
We also display the future direct detection limit set by the XENONnT~\footnote{XENONnT stands for the future limits set by data from XENON 20T$\times$year observations.}.
The red points in the $\mathbb{Z}_2$ breaking scenario are expected to be testable by the future facilities such as XENONnT, PandaX-4T~\cite{Zhang:2018xdp} or LZ~\cite{Akerib:2018lyp}.
We note that in Ref.~\cite{Chiang:2017nmu}, for the $\mathbb{Z}_2$ breaking case studied there, the fixed mass of $h_2$ highly reduces the possibility of achieving a SFOEWPT.


\section{The SFOEWPT, GW signals and the precision test at $e^+e^-$ colliders}
\label{section:SFOEWPT}

\subsection{The finite-temperature effective potential}

In order to evaluate the EWPT in the cxSM, we follow the recipe of Ref.~\cite{Chiang:2017nmu} by using the high-temperature expanded effective potential in order to avoid the gauge dependence problem.
The EWPT is driven by the cubic terms in the effective potential.
Thus, we take the following high-temperature expansion of
\beqs\label{eqs:VT_highT}
\beqn
V(h\,,S\,;T)&=& V_0(h\,,S\,,A=0)+\hf  \Pi_h(T) h^2 + \hf \Pi_S(T) S^2 \,,\\
\Pi_h(T) &=& \Big( \frac{ 2 m_W^2  + m_Z^2 + 2 m_t^2  }{ 4 v^2 } + \frac{\lambda }{2} + \frac{\delta_2}{ 24}   \Big) T^2\,,\\
\Pi_S(T) &=& \frac{1}{12} ( \delta_2 + d_2 ) T^2 \,,
\eeqn
\eeqs
where the finite temperature corrections are given by the thermal mass contributions $\Pi_h(T)$ and $\Pi_S(T)$.

\begin{figure}
\centering
\includegraphics[height=6cm,width=8cm]{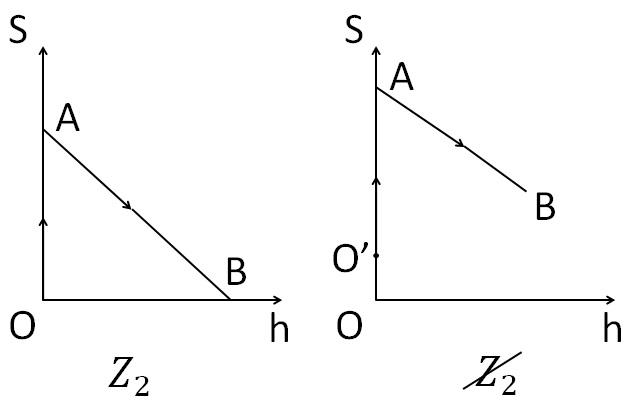}
\caption{The phase transitions for the $\mathbb{Z}_2$ symmetric scenario (left) and the $\mathbb{Z}_2$ breaking scenario (right).
}
\label{fig:EWPT}
\end{figure}

The history of the phase transitions from the high-temperature epoch to the vacuum today is displayed in Fig.~\ref{fig:EWPT} for the $\mathbb{Z}_2$ symmetric scenario (left) and the $\mathbb{Z}_2$ breaking scenario (right).
One can see that the Universe follows a two-step symmetry breaking in both cases in the space of two order parameters for doublet and singlet scalars.
The global minimum of both cases at the high-temperature epoch happens at the spot ``O'' with restored electroweak symmetry.
For the $\mathbb{Z}_2$ symmetric scenario with zero $|a_1|$, when the temperature of the Universe falls down to ``A'', we expect a first minimum with $\langle S \rangle \neq 0$ and $\langle h \rangle =0$.
Along with the further temperature decreasing, a second minimum of with $\langle S \rangle = 0$ and $\langle h \rangle \neq 0$ develops, which eventually becomes the present vacuum at the spot ``B''.
The critical temperature $T_c$ is given when two minima of ``A'' and ``B'' are degenerate.
As there is a barrier between these two minima, a first-order EWPT happens.
For the $\mathbb{Z}_2$ breaking scenario with $a_1\neq 0$, the origin at high temperature is shifted along the $S$ direction from spot $O$ to $O^\prime$ due to the non-vanishing $a_1$ term.
The electroweak symmetry then follows a two-step phase transition process as well.

We find that, with the current potential and thermal correction, only the two-step transition can achieve the strong first-order phase transition.
No thermal barrier can be generated in the one-step transition case as $\langle S \rangle = 0$ in phase ``A''.
The Ref.~\cite{Chiang:2017nmu} included additional thermal corrections and found no parameters leading to the one-step SFOEWPT.
The inclusion of additional cubic $\gU(1)$ breaking terms may exhibit a one-step SFOEWPT.

\subsection{The GW signals}

The GW signals generated during the EWPT depend on the evaluation of the tunneling rate per unit time per unit volume, which is given by~\cite{Moore:1995si}
\beqn
&& \Gamma \sim A(T) \exp(-S_3/T)\,,
\eeqn
with $S_3$ being the Euclidean action of the critical bubble that minimizes the finite-temperature action of
\beqn
S_3&=& 4\pi \int r^2 dr \Big[  \hf \Big(  \frac{d h(r)}{dr} \Big)^2 + \Big(  \frac{d S(r)}{dr} \Big)^2 + V(h\,,S\,,A\,,T)   \Big] \,.
\eeqn
The bubble nucleation temperature $T_n$ is defined as the probability for a single bubble to be nucleated within one horizon volume being $\mO(1)$, that is
\beqn
&& \int_{T_n}^\infty \frac{dT}{T} \Big( \frac{2\zeta M_{\rm pl}}{T}  \Big)^4 \exp(-S_3/T) \sim \mO(1)\,,
\eeqn
where $M_{\rm pl}=1.2\times 10^{19}\,\GeV$ is the Planck mass, and $\zeta \simeq 3\times 10^{-2}$.
Numerically, this equation implies that $S_3(T_n)/T_n \approx 140$ for the EWPT~\cite{Apreda:2001us}.
Two other parameters that are directly relevant to the GW signal calculations are given by
\beqn\label{eq:GWparam}
&& \alpha\equiv \frac{ \rho_{\rm vac} }{\rho_{\rm rad}^*}\, , \quad  \frac{\beta}{H_n} \equiv T_n \frac{ dS_3}{ dT} \Big|_{T_n}\,,
\eeqn
where $\rho_{\rm vac}$ stands for the latent heat released during the EWPT, $H_n$ is the Hubble parameter at $T_n$, and $\rho_{\rm rad}^* = g_* \pi^2 T_n^4/30$ with $g_*$ representing the relativistic degrees of freedom at $T_n$.
Typically, a relatively larger $\alpha$ accompanied with a small $\beta/H_n$ will trigger the SFOEWPT and a significant GW signal.

The observed GW signal is characterized by the energy spectrum $\Omega_{\rm GW}(f)h^2$~\cite{Apreda:2001us}
\beqn
\Omega_{\rm GW}(f)h^2&\equiv& \frac{h^2}{\rho_c} \frac{d \rho_{\rm GW} }{d \log f} \,.
\eeqn
The total energy spectrum here is dominated by the summation of two terms: (1) the sound waves after the bubble collisions; (2) MHD turbulence
\beqn
\Omega_{\rm GW}(f)h^2&\approx&  \Omega_{\rm sw}(f)h^2 + \Omega_{\rm turb}(f)h^2  \,.
\eeqn
The GW signals from the sound wave contribution is given by
\beqn\label{eq:GWsound}
\Omega_{\rm sw}(f) h^2&=& 2.65 \times 10^{-6} \Big( \frac{\beta}{H_n} \Big)^{-1} \Big(  \frac{ \kappa_v \alpha}{ 1+ \alpha}  \Big)^2 \Big(  \frac{100}{g_*} \Big)^{1/3} \non
&& \times v_w \Big(  \frac{f}{f_{\rm sw} } \Big)^3 \Big[  \frac{7}{4 + 3 (f/f_{\rm sw} )^2 }  \Big]^{7/2}\,,
\eeqn
where $\kappa_v$ represents the fraction of latent heat transferred into the bulk motion of the fluid, and was estimated in Ref.~\cite{Caprini:2015zlo}.
The peak frequency $f_{\rm sw}$ is rescaled from its values at the phase transition by
\beqn\label{eq:fsw}
f_{\rm sw}&=& f_{\rm sw}^n \times \frac{a(T_n)}{a_0} = \frac{2\, \beta}{ \sqrt{3}\, v_w } \times \frac{a(T_n)}{a_0}\,.
\eeqn
The MHD turbulence contribution to the GW energy spectrum is written as
\beqn\label{eq:GWturb}
\Omega_{\rm turb}(f) h^2&=& 3.35\times 10^{-4} \Big( \frac{\beta}{ H_n}  \Big)^{-1}  \Big( \frac{\kappa_{\rm tu} \alpha }{ 1+ \alpha}   \Big)^{3/2}  \Big( \frac{100}{g_*}  \Big)^{1/3} \non
&& \times v_w \frac{ (f/f_{\rm tu} )^3 }{ ( 1 +  f/f_{\rm tu}  )^{11/3} \Big( 1+ 8\pi f a_0/ (a(T_n) H_n )  \Big) }\,,
\eeqn
where $\kappa_{\rm tu} \approx 0.1 \kappa_v$.
The peak frequency from the MHD turbulence term is given by
\beqn\label{eq:ftu}
f_{\rm tu}&=& f_{\rm tu}^n  \times \frac{a(T_n)}{a_0}  \approx  \frac{3.5\,\beta}{2\, v_w} \times \frac{a(T_n)}{a_0}\,.
\eeqn
In this study, we consider detonation bubbles and take the wall velocity as a function of $\alpha$ as given in Ref.~\cite{Kamionkowski:1993fg}.
We also note that, to be compatible with EWBG, the wall velocity $v_w$ needs to be obtained as a function of $\alpha$~\cite{Bian:2019zpn,Alves:2018oct,Alves:2018jsw,Alves:2019igs} after taking into account Hydrodynamics.

The discovery prospects of the GW signals are determined by the signal-to-noise ratio (SNR)~\cite{Caprini:2015zlo}
\beqn\label{eq:GWSNR}
{\rm SNR}&=& \sqrt{ \delta\times \mT \int_{f_{\rm min} }^{ f_{\rm max}} df\, \Big[ \frac{ \Omega_{\rm GW}(f) h^2  }{ \Omega_{\rm exp}(f) h^2 }  \Big]^2  }\,,
\eeqn
where $\Omega_{\rm exp}(f) h^2$ stands for the experimental sensitivity for the proposed GW programs.
$\mT$ is the mission duration in years for each experiment, and we assume it to be five here.
The factor $\delta$ counts the number of independent channels for cross-correlated detectors, which is taken to be $1$ for the LISA program~\cite{Thrane:2013oya}.
In practice, we evaluate the SNRs for each benchmark points that achieve the SFOEWPT.
For the LISA program, we take the threshold SNR of $50$ for discovery.
This corresponds to the least sensitive configuration of C4 with four links~\cite{Caprini:2015zlo}\footnote{For a fresh look at GW from first-order phase transition, we referee to Ref.~\cite{Alanne:2019bsm}.}.

\subsection{The precision test at the future $e^+ e^-$ colliders}
\label{section:precision}

\subsubsection{The one-loop corrections to the Higgs boson couplings}

\begin{figure}
\centering
\includegraphics[width=0.6\textwidth,trim=20 100 20 100]{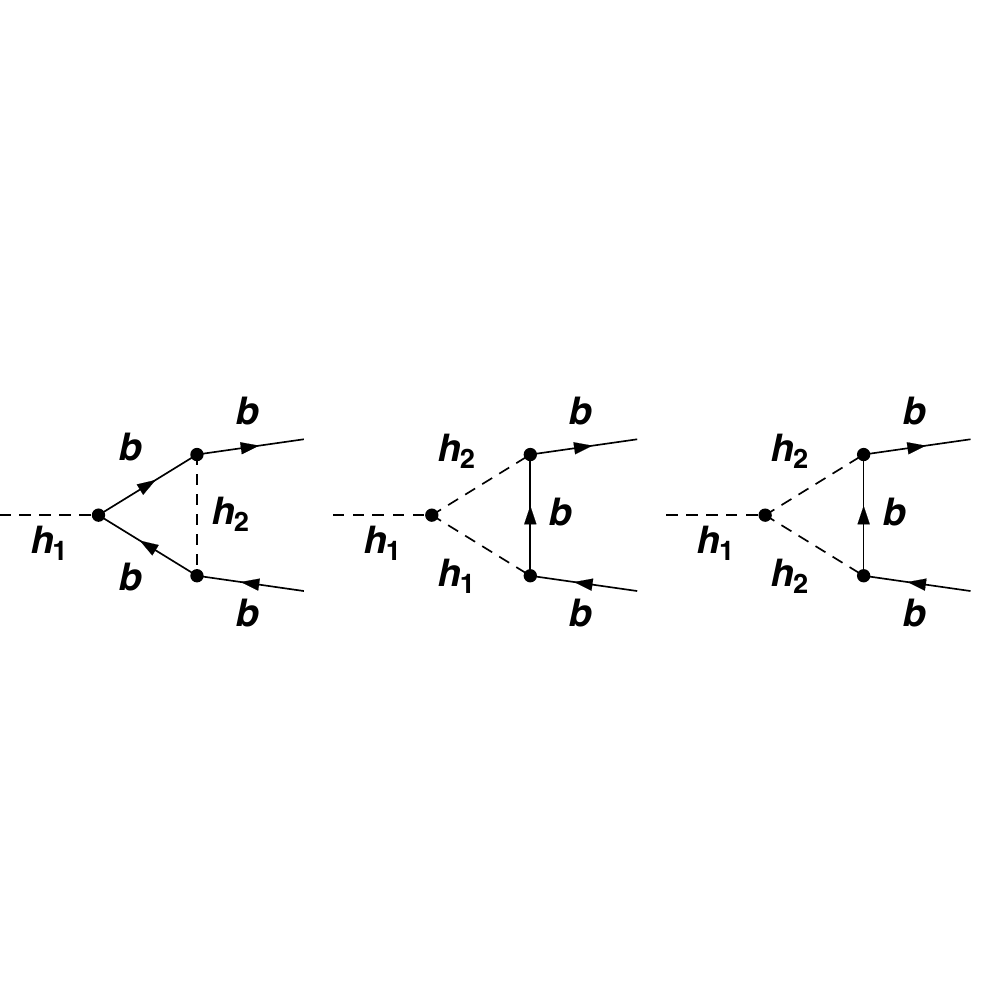}\\
\includegraphics[width=0.6\textwidth,trim=20 100 20 100]{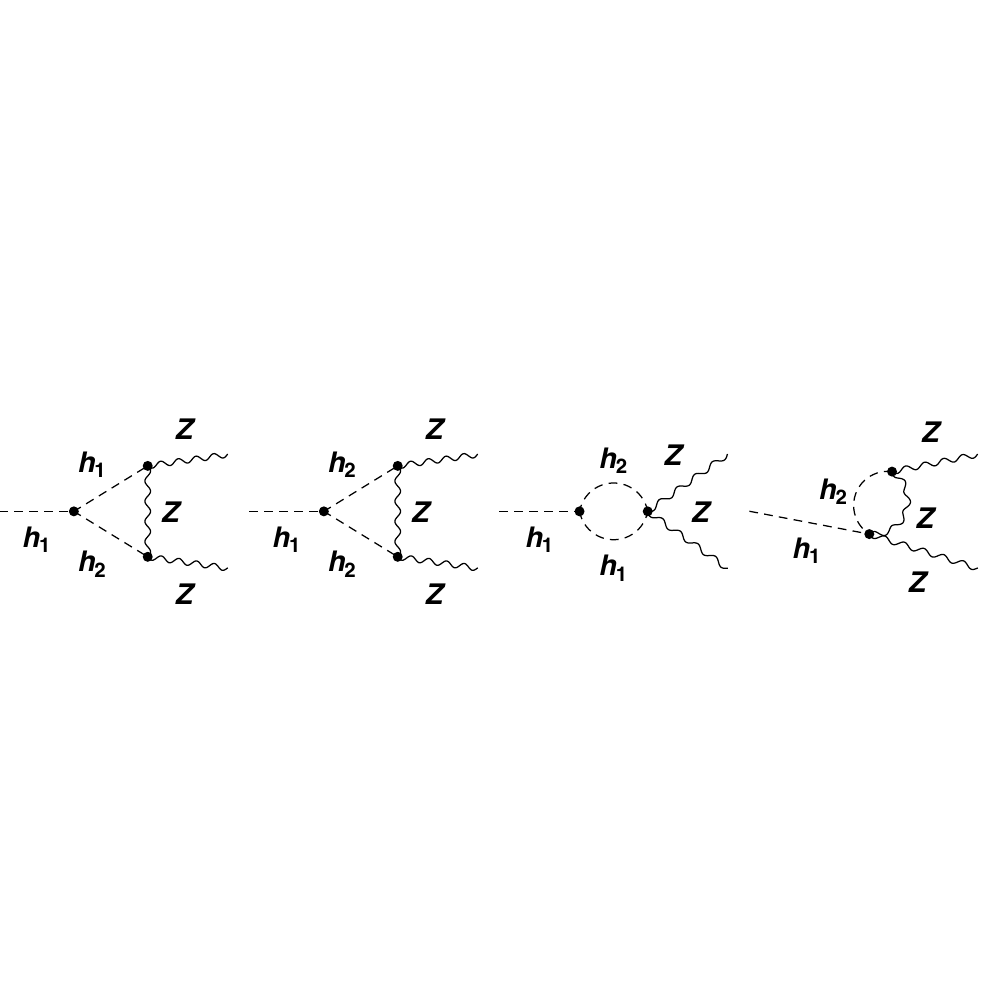}
\caption{
Some representative Feynman diagrams of the one-loop corrections to the SM-like Higgs boson decays $h_1\to b \bar b$ (top) and $h_1 \to ZZ$ (bottom) from the $\mathbb{Z}_2$ breaking scenario in the cxSM.
}
\label{fig:cxSMoneloop}
\end{figure}

The $125\,\GeV$ SM-like Higgs boson can receive corrections from both SM sector as well as the extended scalar sector in the cxSM.
The SM-like Higgs couplings normalized to its SM value, $\kappa$, is defined as~\cite{Chen:2018shg}:
\beqn\label{eq:kappa}
  \kappa_{\rm loop}^{\rm cxSM}&\equiv& \frac{ g_{\rm tree}^{\rm cxSM} + g_{\rm loop}^{\rm cxSM} }{   g_{\rm tree}^{\rm SM} + g_{\rm loop}^{\rm SM} }\, ,
\eeqn
where $g_{\rm tree}^{\rm SM(cxSM)}$ and $g_{\rm loop}^{\rm SM(cxSM)}$ are the couplings in the SM (cxSM) at tree and one-loop level, respectively.
In the $\mathbb{Z}_2$ breaking case, the couplings of SM-like Higgs boson $h_1$ to all SM fields are universally proportional to a factor of $\cos\theta$ due to the singlet-doublet mixing.
The new one-loop contributions~\footnote{In the renormalization scheme we used, the self-energy corrections enter through counter terms, thus we don't show them in the plots.} from cxSM to $h_1 b \bar b$ and $h_1ZZ$ are shown in~\autoref{fig:cxSMoneloop}.
Note that, in the $\mathbb{Z}_2$ symmetric case, although we have $hSS$ and $hAA$ coupling, the contributions in~\autoref{fig:cxSMoneloop} are zero due to the vanishing $S(A)VV$ and $S(A)ff$ couplings.
As a consequence, in the $\mathbb{Z}_2$ symmetric case, the modification of the couplings mainly comes from the Higgs self-energy corrections from $S$ and $A$ loops.
Hence, the deviations of $\kappa$'s in the $\mathbb{Z}_2$ symmetric case are quite universal.
The general vertices of the SM-like Higgs with a pair of gauge bosons $hVV$ and SM fermions $hf\bar{f}$ take the following forms
\begin{subequations}
  \begin{align}
    \Gamma_{hVV}^{\mu\nu}(p_1^2\,,p_2^2\,, q^2)&= \Gamma_{h VV}^1 \eta^{\mu\nu} + \Gamma_{hVV}^2 \frac{ p_1^\mu p_2^\nu }{m_V^2} + i \Gamma_{hVV}^3 \epsilon^{\mu\nu\rho\sigma} \frac{ p_{1\rho} p_{2\sigma} }{ m_V^2} \,,\\
\Gamma_{hf \bar f}(p_1^2\,,p_2^2\,, q^2)&= \Gamma_{h f \bar f}^S + \Gamma_{h f \bar f}^P \gamma_5 + \Gamma_{h f \bar f}^{V_1} \pslash_1 + \Gamma_{h f \bar f}^{V_2}  \pslash_2 \non
&+ \Gamma_{h f \bar f}^{A_1} \pslash_1 \gamma_5 + \Gamma_{h f \bar f}^{A_2}  \pslash_2 \gamma_5 + \Gamma_{h f \bar f}^{T} \pslash_1 \pslash_2 + \Gamma_{h f \bar f}^{PT} \pslash_1 \pslash_2 \gamma_5  \,,
  \end{align}
\end{subequations}
where $(q\,,p_1\,,p_2)$ represent the momenta of the SM-like Higgs boson and two final-state particles.
The $\kappa_i$ for each vertex is given by $\Gamma_{h VV}^1$ and $\Gamma_{h f \bar f}^S$ for the $hVV$ and $h f \bar f$ vertices as
\begin{subequations}
  \begin{align}
    \kappa_V&=\frac{\Gamma_{hVV}^{1}(m_V^2\,,m_h^2\,, q^2)_{\rm cxSM} }{ \Gamma_{hVV}^1( m_V^2\,, m_h^2\,, q^2)_{\rm SM} } \,,\\
\kappa_f&=\frac{\Gamma_{h f \bar f}^{S}(m_f^2\,,m_f^2\,, q^2)_{\rm cxSM} }{ \Gamma_{hf \bar f}^1( m_f^2\,, m_f^2\,, q^2)_{\rm SM} } \,.
  \end{align}
\end{subequations}

In practice, the one-loop corrections to the SM-like Higgs boson couplings are evaluated by adopting the on-shell renormalization scheme~\cite{Denner:1991kt}.
All counter terms, renormalization constants and renormalization conditions are implemented into model files of {\tt FeynArts}~\cite{Hahn:2000kx}, which is then used to generate all possible one-loop diagrams for corresponding couplings for cxSM. After that, {\tt FormCalc}~\cite{Hahn:2016ebn} is used to calculate the full loop level couplings.
The numerical results are performed by {\tt LoopTools}~\cite{Hahn:1998yk}~\footnote{The numerical estimates are summarized in a github repository of \url{https://github.com/ycwu1030/cxSM_Calc}.}.
However, the observables in each experiment are the signal strength $\mu_i$'s instead of $\kappa$'s.
Thus, for each channel, we will calculate the signal strength by
\begin{align}
  \mu_{i\to h\to f} \equiv \frac{\kappa_i^2\kappa_f^2}{\kappa_{\rm width}}\, ,
\end{align}
where $\kappa_{i,f}$ are the normalized coupling relevant for production and decay, and $\kappa_{\rm width}$ represents the ratio of the total width of the SM-like Higgs boson in cxSM to that in SM.

To constrain the model parameters from the current and future Higgs boson precision measurements, a global fit to the observed signal strength is performed with the profile likelihood method.
The $\chi^2$ is defined as
\beqn\label{eq:chi2Higgs}
\chi^2&\equiv& \sum_i  \frac{ ( \mu_i^{\rm cxSM} - \mu_i^{\rm obs} )^2 }{ \sigma_{\mu_i}^2 } \,,
\eeqn
where we sum over all measurements available in experiments and neglect the correlations between different measurements.
In our analyses, $\mu_i^{\rm obs}$ are set to be the SM value $\mu_i^{\rm obs}=1$, for the future colliders.
The estimated errors $\sigma_{\mu_i}$ are listed in~\autoref{tab:mu_precisionCEPCFCC} for the future circular $e^+ e^-$ colliders (CEPC and FCC-ee), and also in~\autoref{tab:mu_precisionILC} for the future ILC program.

\begin{table}[htb]
 \begin{center}
  \begin{tabular}{|l|c|c|c|c|c|}
   \hline
   collider  & \multicolumn{1}{c|}{CEPC}    & \multicolumn{3}{c|}{FCC-ee}   \\
   \hline
   $\sqrt{s}$   &  $\text{240\,GeV}\, hZ $   &  $\text{250\,GeV}\, hZ $  &  $\text{365\,GeV}\, hZ $   &  $\text{365\,GeV}\, h\nu \bar \nu $   \\
   $\int{\mathcal{L}}dt $  &  $\text{5 ab}^{-1} $   &  $\text{5 ab}^{-1} $   &  $\text{ ab}^{-1} $    &  $\text{ ab}^{-1} $  \\
   \hline
    \hline
   $h \to b\bar{b}$    &  0.27\%    &  0.3\%   & 0.5\%   &  0.9\%  \\

   $h \to c\bar{c}$    & 3.3\%    &  2.2\%  & 6.5\%  &  10\% \\

   $h \to gg$    & 1.3\%    &  1.9\% & 3.5\%  & 4.5\%  \\

   $h \to WW^*$   & 1.0\%  & 1.2\%   & 2.6\%  &   3\%  \\

   $h \to \tau^+\tau^-$   & 0.8\%   &  0.9\%  & 1.8\%  &  8\%  \\

   $h \to ZZ^*$    & 5.1\%  &  4.4\%   &  12\% &  10\%   \\

   $h \to \gamma\gamma$   & 6.8\%   & 9.0\%  &  18\% &  22\%    \\

   $h \to \mu^+\mu^-$  & 17\%  &  19\%  & 40\%  & -  \\
  \hline
  \end{tabular}
  \caption{Estimated statistical precisions for Higgs boson measurements obtained at the proposed CEPC~\cite{CEPC-SPPCStudyGroup:2015csa,CEPCStudyGroup:2018ghi} and FCC-ee~\cite{Gomez-Ceballos:2013zzn} programs.
 }
\label{tab:mu_precisionCEPCFCC}
  \end{center}
\end{table}

\begin{table}[htb]
 \begin{center}
  \begin{tabular}{|l|c|c|c|c|c|}
   \hline
   collider   & \multicolumn{5}{c|}{ILC}      \\
   \hline
   $\sqrt{s}$    &  $\text{250\,GeV}\, hZ $   &  $\text{350\,GeV}\, hZ $ &  $\text{350\,GeV}\, h \nu \bar \nu $   &  $\text{500\,GeV}\, hZ $  &  $\text{500\,GeV}\, h\nu\bar \nu $     \\
   $\int{\mathcal{L}}dt $   &  $\text{2 ab}^{-1} $   &  $\text{ ab}^{-1} $  &  $\text{ ab}^{-1} $  &  $\text{ ab}^{-1} $   &  $\text{ ab}^{-1} $      \\
   \hline
    \hline
   $h \to b\bar{b}$     &   0.46\%  &  1.7\%  &  2.0\%  &   0.63\%  &  0.23\%    \\

   $h \to c\bar{c}$     &   2.9\%   &  12.3\%   &  21.2\%  &  4.5\% & 2.2\%   \\

   $h \to gg$      &   2.5\%   &   9.4\%   &  8.6\%  &  3.8\%  &  1.5\%   \\

   $h \to WW^*$    &   1.6\% &   6.3\%  &  6.4\% & 1.9\%  & 0.85\%   \\

   $h \to \tau^+\tau^-$   &  1.1\%  &  4.5\%   & 17.9\%  & 1.5\%  & 2.5\%    \\

   $h \to ZZ^*$    &  6.4\%  & 28.0\%   &  22.4\%  & 8.8\%  &  3.0\%   \\

   $h \to \gamma\gamma$    &  12.0\%  &  43.6\%  & 50.3\%  & 12.0\%  & 6.8\%    \\

   $h \to \mu^+\mu^-$   &  25.5\%  &  97.3\% &  178.9\% &  30\% &  25\%   \\
  \hline
  \end{tabular}
  \caption{Estimated statistical precisions for Higgs boson measurements obtained at the proposed ILC~\cite{Bambade:2019fyw} program.
 }
\label{tab:mu_precisionILC}
  \end{center}
\end{table}

\subsubsection{The electroweak precision tests}

Beside of the SM-like Higgs boson couplings, the model will also change the electroweak observables.
To take this into account, the Peskin-Takuechi parameters~\cite{Peskin:1991sw} of $S$, $T$, and $U$ are used to represent the electroweak precision measurements.
However, $S$ and $A$ have vanishing gauge couplings in the $\mathbb{Z}_2$ symmetric case.
Thus, they do not modify the $S$, $T$ and $U$ parameters.
In the $\mathbb{Z}_2$ breaking case, the expressions for the modifications are given by
\beqs
\beqn
\Delta S &=& \frac{s_\theta^2}{m_Z^2\pi}\Big[ m_Z^2((B_0(m_Z^2,M_1^2,m_Z^2)-B_0(0,M_1^2,m_Z^2))-(B_0(m_Z^2,M_2^2,m_Z^2)-B_0(0,M_2^2,m_Z^2)))\non
&+&  (B_{00}(m_Z^2,M_2^2,m_Z^2)-B_{00}(0,M_2^2,m_Z^2)) \non
&-& (B_{00}(m_Z^2,M_1^2,m_Z^2)-B_{00}(0,M_1^2,m_Z^2)) \Big]\,,\\
\Delta T &=& \frac{s_\theta^2}{4s_W^2m_W^2\pi} \Big[ m_W^2B_0(0,M_1^2,m_W^2)-m_Z^2B_0(0,M_1^2,m_Z^2) - m_W^2B_0(0,M_2^2,m_W^2)+m_Z^2B_0(0,M_2^2,m_Z^2) \non
&+&B_{00}(0,M_1^2,m_Z^2)-B_{00}(0,M_1^2,m_W^2)+B_{00}(0,M_2^2,m_W^2)-B_{00}(0,M_2^2,m_Z^2) \Big]\,,\\
\Delta U &=& -\frac{s_\theta^2}{m_W^2m_Z^2\pi} \Big[ m_W^2m_Z^2(B_0(0,M_1^2,m_W^2)-B_0(0,M_1^2,m_Z^2)-B_0(0,M_2^2,m_W^2)+B_0(0,M_2^2,m_Z^2) \non
&&-B_0(0,m_W^2,M_1^2,m_W^2)+B_0(m_W^2,M_2^2,m_W^2)+B_0(m_Z^2,M_1^2,m_Z^2)-B_0(m_Z^2,M_2^2,m_Z^2))\non
&& +m_Z^2(B_{00}(0,M_2^2,m_W^2)-B_{00}(0,M_1^2,m_W^2)+B_{00}(m_W^2,M_1^2,m_W^2)-B_{00}(m_W^2,M_2^2,m_W^2))\non
&& +m_W^2(B_{00}(0,M_1^2,m_Z^2)-B_{00}(0,M_2^2,m_Z^2)+B_{00}(m_Z^2,M_2^2,m_Z^2)-B_{00}(m_Z^2,M_1^2,m_Z^2))  \Big] \,.
\eeqn
\eeqs
All these modifications are proportional to the CP-even mixing angle of $s_\theta$.
The loop functions of $B_0$ and $B_{00}$ follow the convention in {\tt LoopTools}~\cite{Hahn:1998yk}.

We perform a global fit to the electroweak observables using {\tt Gfitter}~\cite{Baak:2014ora} with current electroweak precisions~\cite{ALEPH:2005ab} and future prospects~\cite{CEPC-SPPCStudyGroup:2015csa,Gomez-Ceballos:2013zzn,Asner:2013psa}.
Unlike the case for Higgs signal strength, the $\chi^2$ constructed from $S$, $T$ and $U$ also includes the correlations among them.
The corresponding $\chi^2$ is thus defined as
\begin{align}
\label{eq:chi2EW}
\chi^2&\equiv \sum_{ij} ( X_i - \hat X_i) (\sigma^2)_{ij}^{-1} ( X_j - \hat X_j)\,,
\end{align}
with $X_i=(\Delta S\,, \Delta T\,, \Delta U)$ being the contributions from the cxSM, and $\hat X_i$ being the corresponding best-fit central values\footnote{For future prospects, the central values are zero.}.
The $\sigma_{ij}^2\equiv \sigma_i \rho_{ij} \sigma_j$ are the error matrix  with uncertainties $\sigma_i$ and correlation matrix $\rho_{ij}$ given in~\autoref{tab:STU} for different experiments. We note that the precision measurements of $W$ boson mass are parameterized by the $S, T$ parameters and are thus embedded in the reach of future $e^+e^-$ colliders in~\autoref{tab:STU}.

\begin{table}[htb]
\centering
\resizebox{\textwidth}{!}{
  \begin{tabular}{|l|c|r|r|r|c|r|r|r|c|r|r|r|c|r|r|r|c|r|r|r|}
   \hline
    & \multicolumn{4}{c|}{Current ($1.7 \times 10^{7}\ Z$'s)}& \multicolumn{4}{c|}{CEPC ($10^{10}Z$'s)}& \multicolumn{4}{c|}{FCC-ee ($7\times 10^{11}Z$'s)}&\multicolumn{4}{c|}{ILC ($10^{9}Z$'s)} \\  \hline
   \multirow{2}{*}{}
   &\multirow{2}{*}{$\sigma$} &\multicolumn{3}{c|}{correlation}
   &{$\sigma$} &\multicolumn{3}{c|}{correlation}
   &{$\sigma$} &\multicolumn{3}{c|}{correlation}
   &{$\sigma$} &\multicolumn{3}{c|}{correlation} \\
   \cline{3-5}\cline{7-9}\cline{11-13}\cline{15-17}
   &&$S$&$T$&$U$&($10^{-2}$)&$S$&$T$&$U$&($10^{-2}$)&$S$&$T$&$U$&($10^{-2}$)&$S$&$T$&$U$\\  \hline
   $S$& $0.04 \pm 0.11$& 1 & 0.92 & $-0.68$ & $2.46$  & 1     & 0.862       & $-0.373$ &   $0.67$    &  1     &   0.812    &    0.001   &   $3.53$    &   1    &    0.988   & $-0.879$ \\ \hline
   $T$&$0.09\pm 0.14$& $-$ & 1 & $-0.87$ & $2.55$  &  $-$   &  1      &  $-0.735$   &   $0.53$    &   $-$    &    1   &    $-0.097$   &    $4.89$   &   $-$    &   1    &   $-0.909$\\  \hline
   $U$& $-0.02 \pm 0.11$& $-$ & $-$ & 1 &$2.08$  &  $-$   &  $-$     &  1   &   $2.40$    &   $-$    &   $-$    &    1   &  $3.76$     &   $-$    &   $-$    & 1 \\ \hline
  \end{tabular}
 }
  \caption{Estimated $S$, $T$, and $U$ ranges and correlation matrices $\rho_{ij}$  from $Z$-pole precision measurements  of the current results, mostly from LEP-I~\cite{ALEPH:2005ab},  and at future lepton colliders CEPC~\cite{CEPC-SPPCStudyGroup:2015csa}, FCC-ee~\cite{Gomez-Ceballos:2013zzn} and ILC ~\cite{Asner:2013psa}.
  {\tt Gfitter} package~\cite{Baak:2014ora} is used in obtaining those constraints.  }
\label{tab:STU}
\end{table}

Besides of the above precision measurements, due to the mixing between Higgs doublet and the scalar singlet in the $\mathbb{Z}_2$ breaking scenario, there exist additional search channels for $h_2$ and corresponding constraints at the LHC.
However, we expect the precision measurements of Higgs and electroweak observables at future colliders place much stronger constraints than the direct search for extra Higgs boson.

\section{Numerical results}
\label{section:results}

Practically, we implement the tree-level effective potential and the high-temperature expansion in Eq.~(\ref{eqs:VT_highT}) into the {\tt CosmoTransitions}~\cite{Wainwright:2011kj}.
The temperature-dependent minima of ``A'' parametrized by $(\langle h \rangle\,, \langle S \rangle) =(0\,, \varphi_S^A)$, and ``B'' parametrized by $(\langle h \rangle\,, \langle S \rangle) =(\varphi_h^B\,, 0)$ for $\mathbb{Z}_2$ symmetric scenario or $(\langle h \rangle\,, \langle S \rangle) =(\varphi_h^B\,, \varphi_S^B)$ for $\mathbb{Z}_2$ breaking scenario are similarly evaluated by using Eq.~\eqref{eqs:VT_highT}.
For the numerical presentation below, we take the data points that not only evade all theoretical constraints and DM constraints, but also achieve the SFOEWPT. The SFOEWPT is characterized by obeying the condition $\varphi_h^B/T_n\equiv v_n/T_n\gtrsim 1$ based on the requirement of the baryon number preservation criterion~\cite{Patel:2011th,Gan:2017mcv,Zhou:2019uzq}.
The {\tt CosmoTransitions}~\cite{Wainwright:2011kj} is used for solving the nucleation temperatures $T_n$, as well as the GW signal parameters of $\alpha$ and $\beta/H_n$.
The solutions of $(T_n\,, \alpha\,, \beta/H_n)$ for each parameter points will be used for the SNRs of the GW signals according to Eq.~\eqref{eq:GWSNR}.
For the precision test of cxSM at future colliders, the results of $\chi^2$ in Eq.~\eqref{eq:chi2Higgs} and Eq.~\eqref{eq:chi2EW} are linearly combined for both Higgs boson and the electroweak precision measurements.
As stated before, we only present the results for $\mathbb{Z}_2$ breaking scenario below as it can lead to a viable DM candidate and meanwhile appears to generate a SFOEWPT.

\subsection{SFOEWPT and GW}

\begin{figure}[htb]
  \centering
  \includegraphics[width=0.45\textwidth]{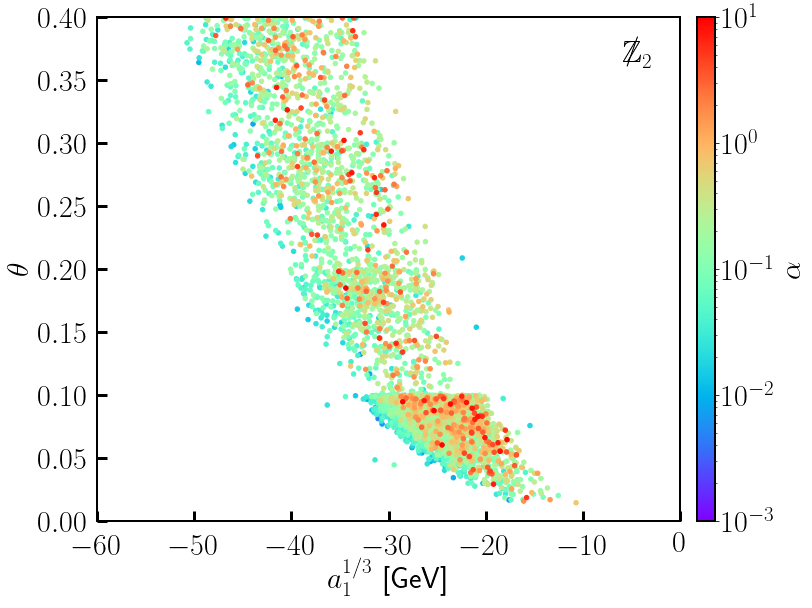}
  \includegraphics[width=0.45\textwidth]{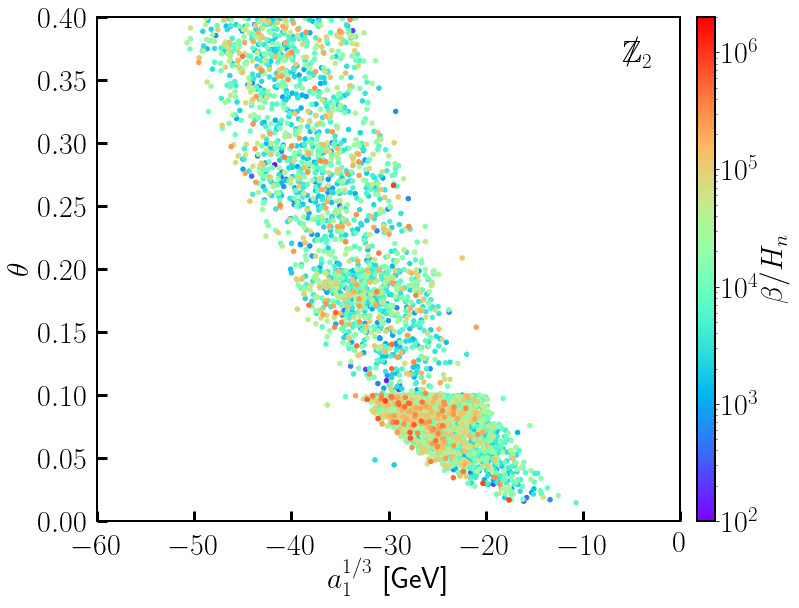}
  \caption{The $\alpha$ (left panels) and $\beta/H_n$ (right panels) in  $a_1^{1/3}$-$\theta$ plane for $\mathbb{Z}_2$ breaking case.  }
  \label{fig:d2_del2_alpha_beta_Z2C_Z2V}
\end{figure}

The GW spectrum is characterized by parameters of $\alpha$ and $\beta/H_n$ defined in Eq.~\eqref{eq:GWparam}, and their values can be fixed by the cxSM potential.
As $\alpha$ and $\beta/H_n$ represent the latent heat released by EWPT and the reversed duration of the EWPT, respectively, significant GW observation typically prefers larger $\alpha$ and smaller $\beta/H_n$ values.
In Fig.~\ref{fig:d2_del2_alpha_beta_Z2C_Z2V}, we display the parameter dependences of $(\alpha\,, \beta/H_n)$ on the cxSM parameters for the $\mathbb{Z}_2$ breaking scenario, with the Higgs and electroweak precision constraints to be imposed later.
The values of $\alpha$ and $\beta/H_n$ are shown in the $(a_1^{1/3}\,,\theta)$ plane.
In the $\mathbb{Z}_2$ breaking case, the minimum at high temperature is shifted by $-a_1$ to positive $S$ direction, which helps achieving the SFOEWPT.
On the other hand, after breaking the global U(1) symmetry, the contribution of $a_1$ term to the mass square of Goldstone boson $A$ is given by $-\sqrt{2}a_1/v_s$. Thus, negative $a_1$ values are preferred as shown in the plots.
One can see that relatively larger $\alpha$ and smaller $\beta/H_n$ values prefer to reside around the region with small $|a_1^{1/3}|$.
It turns out that the shift of the high-temperature minimum by $-a_1^{1/3}$ along the $\varphi_S$ direction should not be rather sizable to break the discrete symmetry in this case.
Although we present all date points satisfying the SFOEWPT criterion here, one should note that those points with $\beta/H_n\gtrsim 10^4$ produce too large peak frequencies as from Eqs.~\eqref{eq:fsw} \eqref{eq:ftu}, and too small power spectrum as from Eqs.~\eqref{eq:GWsound} \eqref{eq:GWturb}.
Thus, such points are impossible to be detected by the GW detectors which are mostly sensitive to milliHz frequencies.

\begin{figure}[!tbp]
  \centering
  \includegraphics[width=0.6\textwidth]{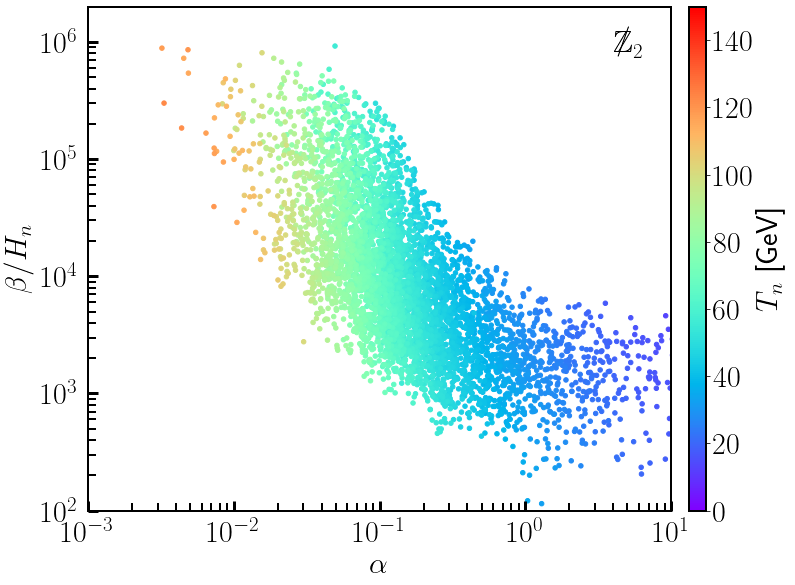}
  \caption{The relationship between $T_n$ and $\alpha$, $\beta/H_n$ parameters for the $\mathbb{Z}_2$ breaking case.
  }
  \label{fig:alpha_beta_Tn}
\end{figure}

In Fig.~\ref{fig:alpha_beta_Tn}, we show the bubble nucleation temperature $T_n$ in the GW parameter plane of $(\alpha\,, \beta/H_n)$.
The lower $T_n$ one obtains, the stronger the EWPT becomes.
In principle, as a result, we can have increased $\alpha$ and decreased $\beta/H_n$.
The realistic situation of their relationship might be more complicated to achieve the bubble nucleation condition while comparing different specific models.
For instance, for the $\mathbb{Z}_2$ breaking case, the origin is shifted at the high temperature.
As seen in the plot, the $\mathbb{Z}_2$ breaking scenario exhibits lower $T_n$ and larger $\alpha$ as a result of a two-step bubble nucleation. The values of $\beta/H_n$ parameter span a broad space and can be relatively large for the $\mathbb{Z}_2$ breaking case.

\begin{figure}[!tbp]
  \centering
  \includegraphics[width=\textwidth]{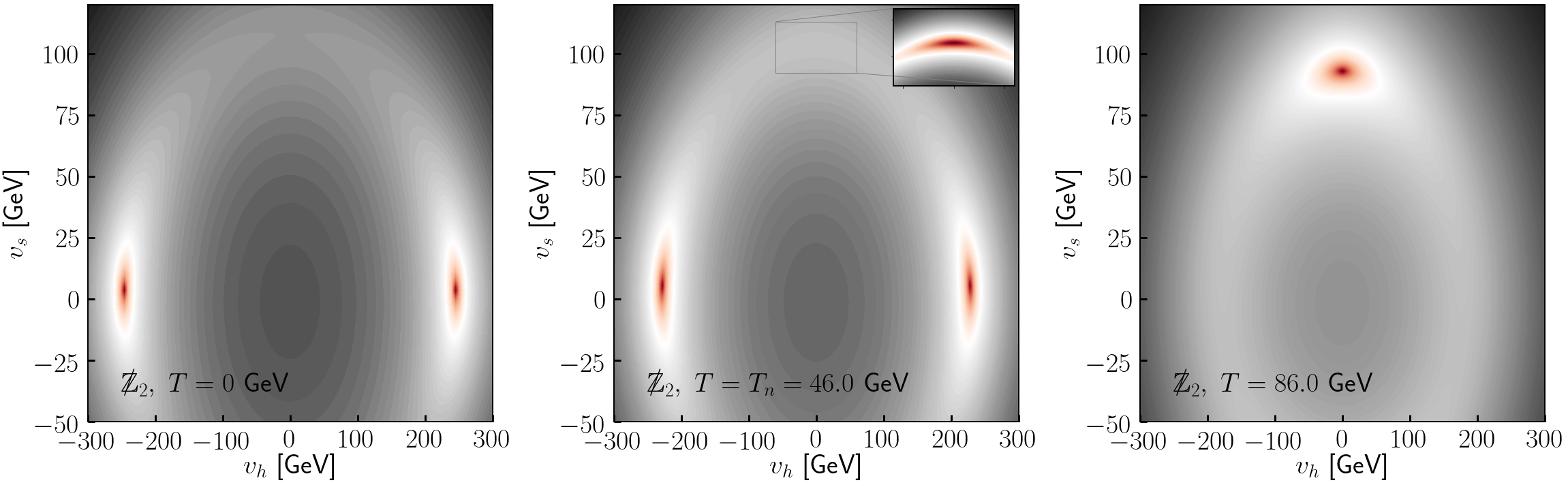}
  \caption{
  The potential for the benchmark point A from $\slashed{\mathbb{Z}}_2$ case at different temperatures (left: $T=0$, middle: $T=T_n$, right: $T=T_n+40$ GeV). In the middle panel, we add zoomed-in windows to clearly indicate that the point is actually a local minimum.
  }
  \label{fig:potential_BM}
\end{figure}

\begin{figure}[!tbp]
  \centering
  \includegraphics[width=0.6\textwidth]{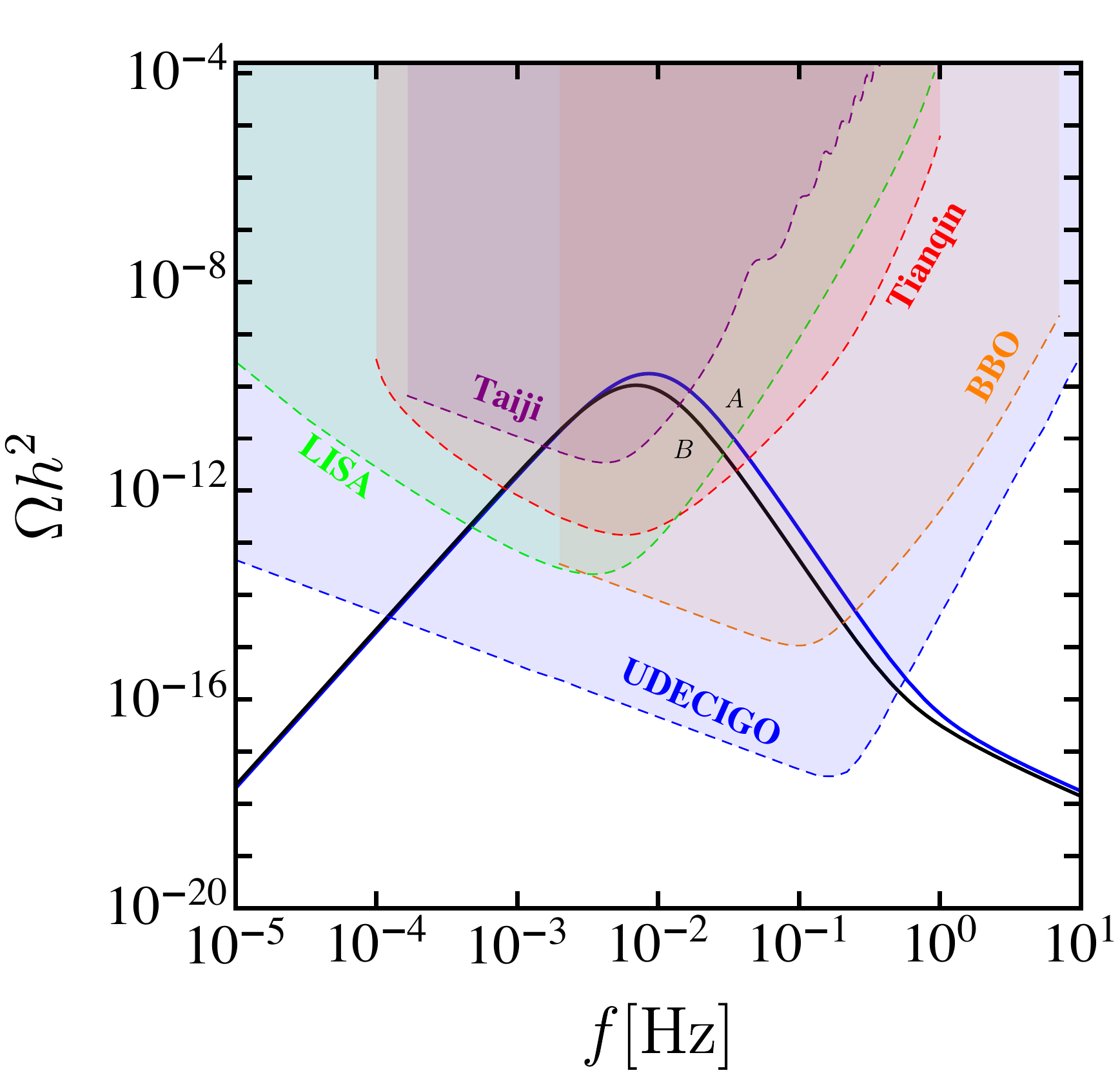}
  \caption{
  The GW spectrum for two benchmark points from the $\slashed{\mathbb{Z}}_2$ case.
  }
  \label{fig:GW_BM}
\end{figure}

\begin{table}[htb]
 \centering{
   \resizebox{\textwidth}{!}{
  \begin{tabular}{ccccccccccccc}
  \hline\hline
  $\slashed{\mathbb{Z}}_2$ & $M_{2}\, [\GeV]$  & $M_A\, [\GeV]$ & $v_s\, [\GeV]$ & $(a_1)^{1/3}\, [\GeV]$ & HL-LHC & $e^+ e^-$ &  SNR (LISA) & $\alpha$ & $\beta/H_n$ & $T_n$ & $v_n/T_n$ & $v_w$ \\
  \hline
  A & $99$ & $963$ & 3.6 & -29.3 & $\times$ & $\surd$  & $3.3\times 10^6$ & $0.31$	& $834.64$ & $46.02$	& $4.96$ & $0.86$\\
   B & $98$ & $984$ & 1.6 & -22.0 & $\times$ & $\times$  & $2.4\times 10^6$ & $0.24$	& $562.59$ & $54.17$	& $4.13$ & $0.84$ \\
  \hline\hline
  \end{tabular}}}
  \caption{
  Two benchmark points for the $\mathbb{Z}_2$ breaking case.
  The marks of $\times$ ($\surd$) represent whether the benchmark point cannnot (can) be probed for the given precision of the corresponding collider runs.
  }
\label{tab:BM_GW}
\end{table}

To compare the GW signal spectra, we list two benchmark points for the $\mathbb{Z}_2$ breaking case in Tab.~\ref{tab:BM_GW}.
The CP-odd scalar masses of these two benchmark points are close to each other, i.e. $M_A=963\,\GeV$ (benchmark point A) and $M_A=984\,\GeV$ (benchmark point B) respectively.
The benchmark point B cannot be searched for via the precision measurements of the Higgs boson decays at the future HL-LHC or $e^+ e^-$ colliders, while it yields a SNR of $\mO(10^6)$ at the LISA.
The benchmark point A can be probed via both the precision measurements of the Higgs boson decays at the future $e^+ e^-$ colliders and the GW spectrum at the LISA, with an SNR of $\mO(10^6)$.
In \autoref{fig:potential_BM}, we show the high-temperature effective potential for the benchmark point A.
One can see that a local minimum, denoted by the red contours, first develops along the $S$ direction at high temperature.
As the temperature cools down, a second minimum occurs along the electroweak symmetry breaking direction, which will become the global minimum latter.
At $T_n$ there exists a barrier between the two minima.
The Universe then tunnels to this global minimum resulting in a SFOEWPT.
The local minimum along the singlet direction becomes saddle point at zero temperature.
This is true for all points in the $\mathbb{Z}_2$ breaking case.
Their GW spectra $\Omega h^2$ versus the frequency $f$ are displayed in Fig.~\ref{fig:GW_BM}, together with the viable signal regions of different ongoing/upcoming GW detection programs.
The benchmark point B in the $\mathbb{Z}_2$ breaking case exhibits higher $T_n$ and corresponding decreased $\alpha$.
The value of $\beta/H_n$ in benchmark point A is larger than that in benchmark point B, which leads to a bit larger peak of frequency in the $\mathbb{Z}_2$ breaking case, together with a lower $T_n$.

\subsection{The precision tests at the colliders}

\begin{figure}[!tbp]
  \centering
  \includegraphics[width=0.3\textwidth]{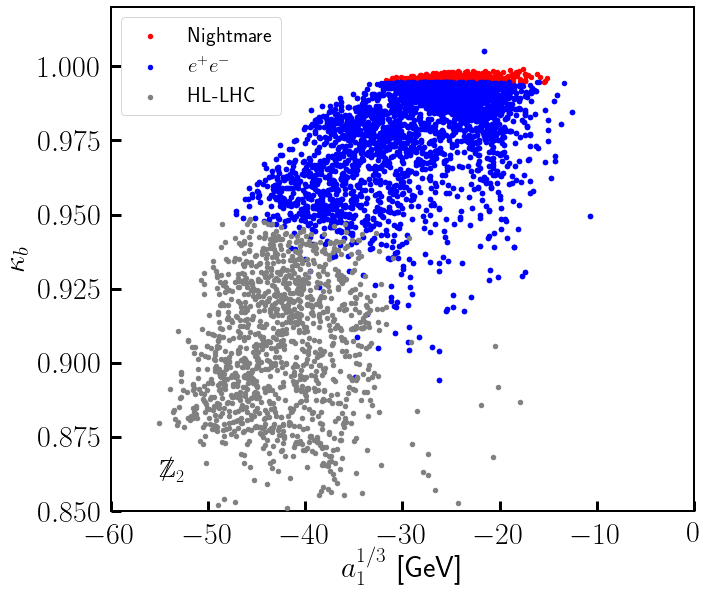}
  \includegraphics[width=0.3\textwidth]{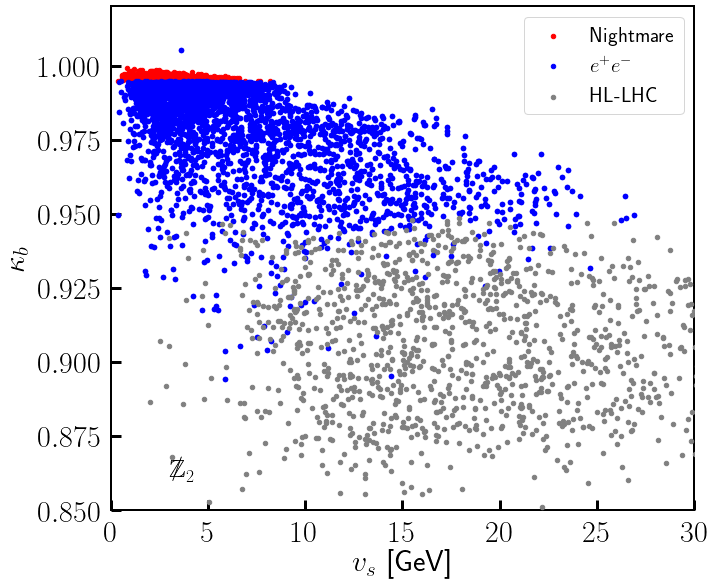}
  \includegraphics[width=0.3\textwidth]{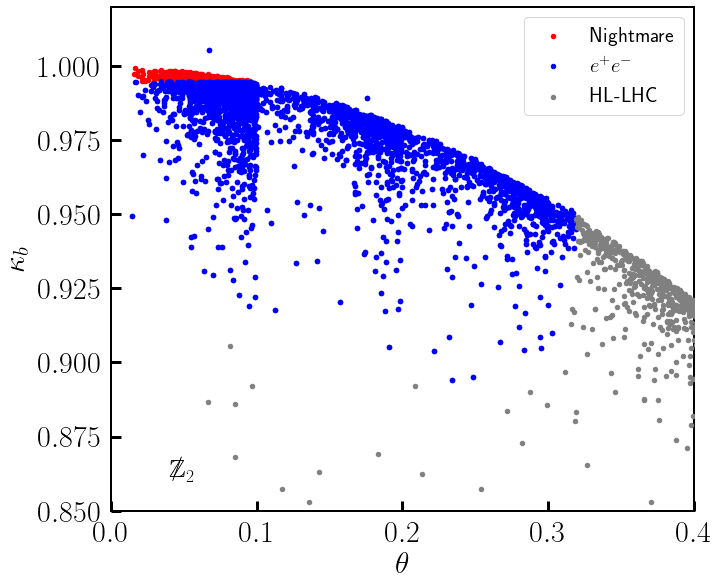}\\
  \includegraphics[width=0.3\textwidth]{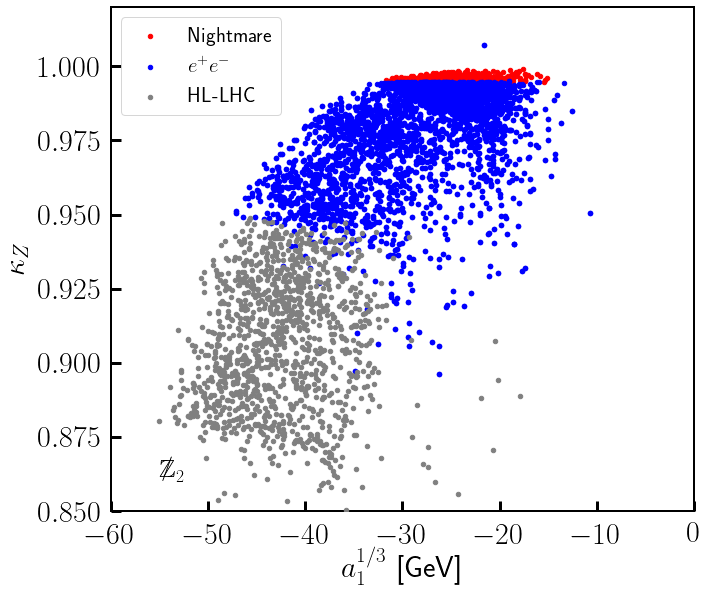}
  \includegraphics[width=0.3\textwidth]{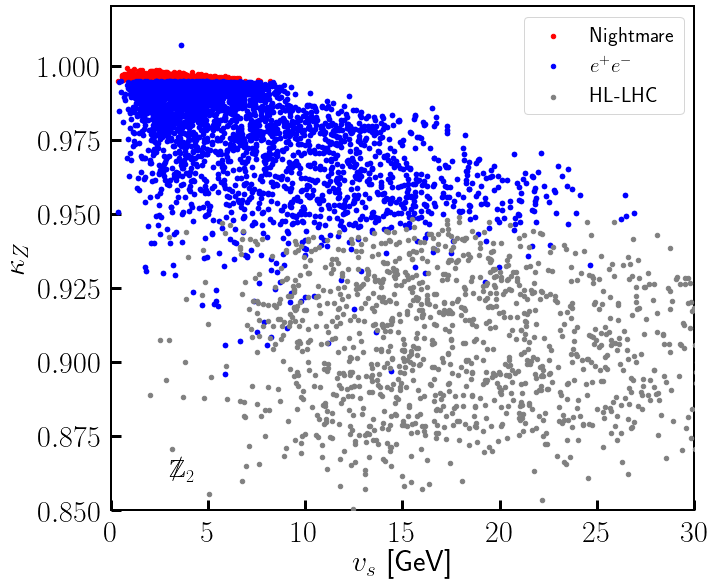}
  \includegraphics[width=0.3\textwidth]{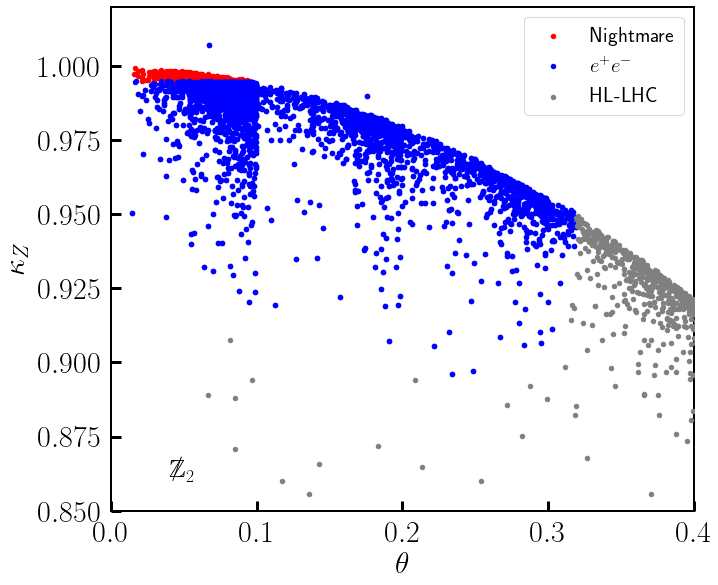}
  \caption{
  The normalized SM-like Higgs boson couplings of $\kappa_b$ (upper panels) and $\kappa_Z$ (lower panels) for the $\mathbb{Z}_2$ breaking scenario.
  The grey points are within the sensitivities of both the HL-LHC and any of the future $e^+e^-$ colliders (including CEPC, FCC-ee, and ILC). Blue points are only within the sensitivities of $e^+e^-$ colliders. The red ones are ``Nightmare'' model points.
  }
  \label{fig:kb_kZ_Z2V}
\end{figure}

The precision tests are made by the combined $\chi^2$ fit of the SM-like Higgs boson measurements and the electroweak precision measurements according to Eqs.~\eqref{eq:chi2Higgs} and~\eqref{eq:chi2EW}.
In Fig.~\ref{fig:kb_kZ_Z2V}, we display two couplings of $\kappa_b$ and $\kappa_Z$ for the $\mathbb{Z}_2$ breaking case.
For the $\mathbb{Z}_2$ breaking case, the normalized Higgs boson couplings are displayed versus the physical parameters of $(a_1^{1/3}\,, v_s\,,\theta)$.
The model points in grey are within the sensitivities of both the HL-LHC and any of the future $e^+e^-$ colliders (including CEPC, FCC-ee and ILC), while the blue points are only within the sensitivities of $e^+e^-$ colliders.
The red points are those cannot be probed by the HL-LHC and future $e^+ e^-$ colliders.
We denote them as ``nightmare'' model points~\cite{Curtin:2014jma,Huang:2016cjm}.
In the $\mathbb{Z}_2$ breaking scenario, there can be sizable one-loop corrections to the SM-like Higgs boson decays through the cxSM sector, as were previously shown in Fig.~\ref{fig:cxSMoneloop}, besides the tree-level correction given by $g_{\rm tree}^{\rm cxSM}/g_{\rm tree}^{\rm SM}=\cos\theta$.
Thus, as seen in Fig.~\ref{fig:kb_kZ_Z2V}, some points can be probe for the search sensitivities of the HL-LHC while some other points with larger $\kappa$ couplings can be probed by future $e^+ e^-$ colliders.

\begin{figure}[htb]
  \centering
  \includegraphics[width=0.45\textwidth]{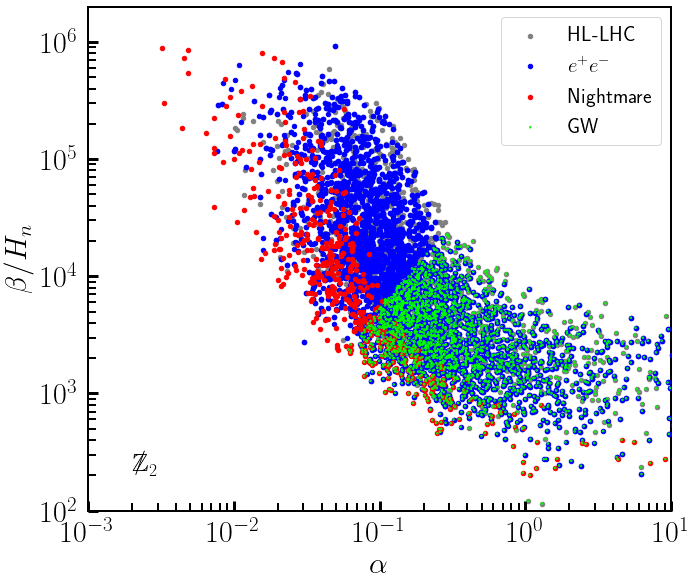}
  \caption{
  The Higgs precision measurement and GW signals for the $\mathbb{Z}_2$ breaking case.
  The grey points (most are hidden under blue points) are those can be probed for the sensitivities of both the HL-LHC and any $e^+e^-$ colliders.
  Blue points are those can only be probed at any of the future electron-positron colliders.
  The red points (``nightmare'') are those that cannot be probed by future colliders.
  The green points are those with SNR$>50$ of the GW signals for the future LISA interferometer.
  }
  \label{fig:alpha_beta_Precision_Z2V_Z2C}
\end{figure}

In Fig.~\ref{fig:alpha_beta_Precision_Z2V_Z2C}, we combine the experimental sensitivities of the colliders and the GW signal probes via the LISA interferometer in the $(\alpha\,, \beta/H_n)$ plane.
For the $\mathbb{Z}_2$ breaking case, the corrections of treel-level and one-loop effects from the cxSM sector become significant.
Correspondingly, we found a majority of model points can be probe for the search sensitivities of both the HL-LHC and the future $e^+ e^-$ runs.
The LISA interferometer is likely to probe the model points with relatively large values of $\alpha$ and small values of $\beta/H_n$.
A smaller fraction of model points (denoted in red) are beyond the search limits by either the Higgs measurements at the future colliders or the LISA interferometer. Nevertheless, these points are all within the sensitivity of future DM DD experiments.

\begin{figure}[!tbp]
  \centering
  \includegraphics[width=0.6\textwidth]{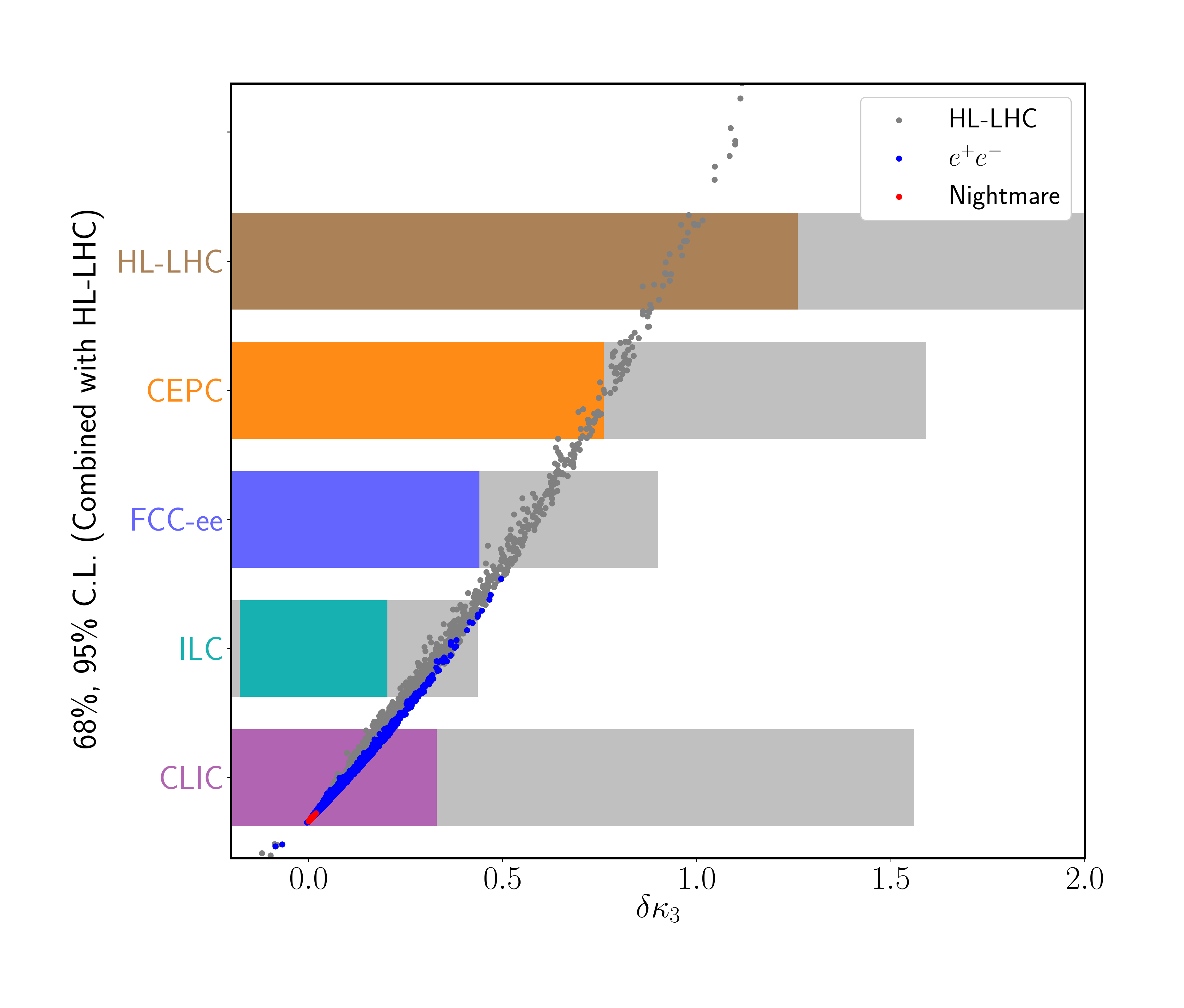}
  \caption{
  The precision measurement of $\delta\kappa_3$ in the $\mathbb{Z}_2$ breaking case from different collider experiments.
  The vertical position of the points is irrelevant.
  The colored and grey shaded regions correspond to 68\% and 95\% C.L. regions, respectively, for $\delta\kappa_3$ from Ref.~\cite{DiVita:2017vrr}.}\label{fig:dk3_dk4}
\end{figure}

Finally, in Fig.~\ref{fig:dk3_dk4}, we show the expected sensitivities of future colliders to $\delta \kappa_3$ in the $\mathbb{Z}_2$ breaking case.
To the right of the colored (grey) bars, the corresponding colliders are sensitive to the measurement of cubic Higgs coupling at 68\% (95\%) C.L.\,.
Note that, in the $\mathbb{Z}_2$ symmetric case, the cubic and quartic Higgs couplings are the same as those in SM.
Thus, we do not have any sensitivity in these measurements. While, in the $\mathbb{Z}_2$ breaking case, the Higgs self-couplings do differ from the SM as shown in Eqs.~(\ref{equ:Z2VCoupling}). The sensitivity of future colliders to the cubic coupling, however, is much lower than the precision measurements of other couplings. We find that, with theoretical constraint and DM constraint, a majority of $\delta \kappa_3$ values are positive and the negative $\delta \kappa_3$ cases in Fig.~\ref{fig:dk3_dk4} correspond to a few points with DM mass near $h_1$ resonance, i.e. $M_A\gtrsim M_1/2$. These points have relatively large $|a_1|$ and/or $\theta$ as shown in Eq.~(\ref{equ:Z2VCoupling}a). The resonant decrease of the DM relic density guarantees the evasion of DD constraint in spite of a large $|a_1|$. However, these points can be tested by future Higgs precision measurements.


\section{Conclusion}
\label{section:conclusion}

In this work, we study the future experimental tests of the cxSM.
Future experimental facilities at the high intensity/energy frontiers, such as GW detection and $e^+e^-$ colliders, can test the visible parameter space of this complex scalar model achieving a SFOEWPT and providing scalar DM candidate.
We apply theoretical constraints and DM constraints from relic density and the latest XENON1T limit on the parameter space of the cxSM, and also require they pass the SFOEWPT criterion.
By combining the $\chi^2$ fit of the SM-like Higgs boson measurements and the electroweak precision measurements, we estimate whether the model points can be accessible at the future $e^+ e^-$ colliders.

In the $\mathbb{Z}_2$ symmetric scenario, the complex scalar singlet $\mathbb{S}$ does not develop a vev and a quadratic term of $\mathbb{S}$ is introduced to break a global U$(1)$ symmetry.
As a result, the real part of $\mathbb{S}$ does not mix with the SM Higgs and both the real and imaginary parts become the DM candidates.
This scenario admits a two-step SFOEWPT in the way that the scalar singlet acquires a vev at the high temperature prior to the electroweak symmetry breaking.
We find that none of the generated model points can achieve a SFOEWPT and meanwhile provide a viable DM candidate.
In this scenairo, as $\mathbb{S}$ does not mix with the SM Higgs doublet, there is no tree-level correction to the SM-like Higgs couplings and the one-loop corrections from the Higgs boson self-energy terms are very small.
We also expect the model points cannot be probed given the sensitivities of the HL-LHC and any future $e^+e^-$ colliders.

In the $\mathbb{Z}_2$ breaking scenario, the complex scalar singlet $\mathbb{S}$ develops a non-vanishing vev and an additional linear term of $\mathbb{S}$ is introduced to break the discrete $\mathbb{Z}_2$ symmetry.
Thus, besides the sizable loop corrections, the mixing of the complex singlet and the SM Higgs doublet induces a tree-level correction to the SM-like Higgs couplings which is the cosine of the mixing angle.
The CP-odd component of the complex singlet serves as the DM candidate.
This scenario also achieves a two-step SFOEWPT driven at tree-level.
It turns out that a majority of model points can be covered by the precision Higgs measurements at the future colliders.

We also find that, in both $\mathbb{Z}_2$ symmetric and $\mathbb{Z}_2$ breaking scenarios, some of the points without the sensitivities of future colliders are accompanied with sizable signal-to-noise ratio around $f\sim \mO( 10^{-4})- \mO(1) \,{\rm Hz}$ for their GW signals.
Future space-based GW interferometer, such as LISA, can thus probe such ``nightmare'' parameter space.
In addition, all the model points realizing the SFOEWPT and satisfying DM constraints are within the sensitivity of future DM DD experiments.


\section*{Acknowledgments}

We would like to thank Yanwen Liu, Guy David Moore, Shi Pi, Michael Ramsey-Musolf, and Lian-Tao Wang for very useful discussions and communication.
NC is partially supported by the National Natural Science Foundation of China (under Grant No. 11575176).
TL is supported by the National Natural Science Foundation of China (Grant No. 11975129) and ``the Fundamental Research Funds for the Central Universities'', Nankai University (Grants No. 63191522, 63196013).
YCW is partially supported by the Natural Sciences and Engineering Research Council of Canada.
LGB is supported by the National Natural Science Foundation of China (under grant No.11605016 and No.11647307).

\newpage


\bibliography{references}

\end{document}